\documentclass[seceqn]{elsart}
\pdfoutput=1

\usepackage{ifpdf}
\usepackage[intlimits]{amsmath}
\usepackage{amsfonts}
\usepackage{amssymb}
\usepackage{mathrsfs}
\ifpdf
  \usepackage[pdftex]{graphicx}
\else
  \usepackage[dvips]{graphicx}
\fi
\usepackage{color}
\usepackage{amssymb}

\journal{Computer Physics Communications}

\begin{document}

\numberwithin{figure}{section}

\newcommand{\del}[2]{\ensuremath{\frac{\partial #1}{\partial #2}}}     
\newcommand{\di}[1]{\ensuremath{~\mathsf{d}#1}}                        
\newcommand{\dd}[2]{\ensuremath{\frac{\mathsf{d} #1}{\mathsf{d}\,#2}}} 
\newcommand{\ddelta}[2]{\ensuremath{\frac{\delta #1}{\delta #2}}}      
\newcommand{\ttfrac}[2]{\ensuremath{\genfrac{}{}{0.02ex}{3}{#1}{\rule{0pt}{0.9ex}#2}}}
\newcommand{\inv}[1]{\ensuremath{\frac{1}{#1}}}                        
\newcommand{\tinv}[1]{\ensuremath{\tfrac{1}{#1}}}                      
\newcommand{\thalf}{\ensuremath{\genfrac{}{}{0.02ex}{3}{1}{\rule{0pt}{0.9ex}2}}}
\newcommand{\vek}[1]{\ensuremath{{\mathbf{#1}}}}                       
\newcommand{\imean}[1]{\ensuremath{\left\langle{#1}\right\rangle}}     
\newcommand{\order}[2]{\ensuremath{\mathcal{O}\left(#1^{#2}\right)}}   
\newcommand{\tv}{\ensuremath{\mathsf{TV}}}                             
\newcommand{\minmod}[2]{\ensuremath{\mathrm{minmod}\left(#1,#2\right)}}
\newcommand{\superbee}[2]{\ensuremath{\mathrm{superbee}\left(#1,#2\right)}}
\newcommand{\mc}[3]{\ensuremath{\mathrm{mc}\left(#1,#2,#3\right)}}     
\newcommand{\cfl}{\ensuremath{\mathsf{CFL}}}                           

\newcommand{\delxi}[1]{\ensuremath{\del{#1}{\xi}}}
\newcommand{\deleta}[1]{\ensuremath{\del{#1}{\eta}}}
\newcommand{\delphi}[1]{\ensuremath{\del{#1}{\phi}}}
\newcommand{\Dxi}{\ensuremath{\mathcal{D}_{\xi}}}
\newcommand{\Deta}{\ensuremath{\mathcal{D}_{\eta}}}

\newcommand{\den}{\ensuremath{\varrho}}             
\newcommand{\Sden}{\ensuremath{\Sigma}}             
\newcommand{\pre}{\ensuremath{p}}                   
\newcommand{\sam}{\ensuremath{\ell}}                
\newcommand{\cs}{\ensuremath{c_s}}                  
\newcommand{\ien}{\ensuremath{\varepsilon}}         
\newcommand{\etot}{\ensuremath{E}}                  
\newcommand{\emod}{\ensuremath{\tilde{e}}}          
\newcommand{\ethp}{\ensuremath{w}}                  
\newcommand{\vel}{\ensuremath{{\vek{v}}}}           
\newcommand{\Cp}{\ensuremath{c_{P}}}                
\newcommand{\Cv}{\ensuremath{c_{V}}}                
\newcommand{\tff}{\ensuremath{t_{\mathrm{ff}}}}     
\newcommand{\tdyn}{\ensuremath{t_{\mathrm{dyn}}}}   
\newcommand{\Teff}{\ensuremath{T_{\mathrm{eff}}}}   
\newcommand{\rbondi}{r_{\mathsf{b}}}                
\newcommand{\msun}{\ensuremath{\mathrm{M}_{\odot}}} 
\newcommand{\Ldisk}{\ensuremath{L_{\mathrm{D}}}}    
\newcommand{\LEdd}{\ensuremath{L_{\mathrm{Edd}}}}   
\newcommand{\se}{\ensuremath{\sigma_{\mathrm{e}}}}  
\newcommand{\vth}{\ensuremath{v_{\mathrm{th}}}}     
\newcommand{\jahr}{\ensuremath{\mathrm{a}}}         
\newcommand{\Rs}{\ensuremath{R_{\mathrm{s}}}}       

\newcommand{\vxi}{\ensuremath{v_{\hat{\xi}}}}                            
\newcommand{\veta}{\ensuremath{v_{\hat{\eta}}}}
\newcommand{\vphi}{\ensuremath{v_{\hat{\phi}}}}
\newcommand{\vxisq}{\ensuremath{v^2_{\hat{\xi}}}}                        
\newcommand{\vetasq}{\ensuremath{v^2_{\hat{\eta}}}}
\newcommand{\vphisq}{\ensuremath{v^2_{\hat{\phi}}}}
\newcommand{\fxi}{\ensuremath{f_{\hat{\xi}}}}                            
\newcommand{\feta}{\ensuremath{f_{\hat{\eta}}}}
\newcommand{\fphi}{\ensuremath{f_{\hat{\phi}}}}
\newcommand{\fz}[1]{\ensuremath{f_{z,\hat{#1}}}}                         
\newcommand{\erot}{\ensuremath{\frac{\den}{2}\vphisq}}                   

\newcommand{\hxi}{\ensuremath{h_{\xi}}}                                  
\newcommand{\heta}{\ensuremath{h_{\eta}}}
\newcommand{\hphi}{\ensuremath{h_{\phi}}}
\newcommand{\dxi}{\ensuremath{\di{\xi}}}                                 
\newcommand{\deta}{\ensuremath{\di{\eta}}}
\newcommand{\dphi}{\ensuremath{\di{\phi}}}
\newcommand{\exi}{\ensuremath{\widehat{e}_{\hat{\xi}}}}                  
\newcommand{\eeta}{\ensuremath{\widehat{e}_{\hat{\eta}}}}
\newcommand{\ephi}{\ensuremath{\widehat{e}_{\hat{\phi}}}}
\newcommand{\cxyx}{\ensuremath{c_{\hat{\xi}\hat{\eta}\hat{\xi}}}}
\newcommand{\cxyy}{\ensuremath{c_{\hat{\xi}\hat{\eta}\hat{\eta}}}}       
\newcommand{\cxzx}{\ensuremath{c_{\hat{\xi}\hat{\phi}\hat{\xi}}}}
\newcommand{\cxzz}{\ensuremath{c_{\hat{\xi}\hat{\phi}\hat{\phi}}}}
\newcommand{\cyxy}{\ensuremath{c_{\hat{\eta}\hat{\xi}\hat{\eta}}}}
\newcommand{\cyxx}{\ensuremath{c_{\hat{\eta}\hat{\xi}\hat{\xi}}}}
\newcommand{\cyzy}{\ensuremath{c_{\hat{\eta}\hat{\phi}\hat{\eta}}}}
\newcommand{\cyzz}{\ensuremath{c_{\hat{\eta}\hat{\phi}\hat{\phi}}}}
\newcommand{\czxz}{\ensuremath{c_{\hat{\phi}\hat{\xi}\hat{\phi}}}}
\newcommand{\czxx}{\ensuremath{c_{\hat{\phi}\hat{\xi}\hat{\xi}}}}
\newcommand{\czyz}{\ensuremath{c_{\hat{\phi}\hat{\eta}\hat{\phi}}}}
\newcommand{\czyy}{\ensuremath{c_{\hat{\phi}\hat{\eta}\hat{\eta}}}}
\newcommand{\Txx}{\ensuremath{T_{\hat{\xi}\hat{\xi}}}}
\newcommand{\Txy}{\ensuremath{T_{\hat{\xi}\hat{\eta}}}}
\newcommand{\Txz}{\ensuremath{T_{\hat{\xi}\hat{\phi}}}}
\newcommand{\Tyx}{\ensuremath{T_{\hat{\eta}\hat{\xi}}}}
\newcommand{\Tyy}{\ensuremath{T_{\hat{\eta}\hat{\eta}}}}
\newcommand{\Tyz}{\ensuremath{T_{\hat{\eta}\hat{\phi}}}}
\newcommand{\Tzx}{\ensuremath{T_{\hat{\phi}\hat{\xi}}}}
\newcommand{\Tzy}{\ensuremath{T_{\hat{\phi}\hat{\eta}}}}
\newcommand{\Tzz}{\ensuremath{T_{\hat{\phi}\hat{\phi}}}}
\newcommand{\DeltaV}[2]{\ensuremath{\Delta V_{#1}^{#2}}}                 

\newcommand{\jaco}[1]{\ensuremath{\biggl(\del{\vek{#1}}{\vek{u}}\biggr)}}
\newcommand{\eigl}[2]{\ensuremath{\vek{l}_{#1}^{(#2)}}}
\newcommand{\eigr}[2]{\ensuremath{\vek{r}_{#1}^{(#2)}}}
\newcommand{\cvar}[2]{\ensuremath{w^{(#1)}_{#2}}}

\newcommand{\alldirections}[3]{\ensuremath{#1_{#2}^{#3}}}
\newcommand{\west}[2]{\alldirections{#1}{#2}{\mathsf{w}}}
\newcommand{\east}[2]{\alldirections{#1}{#2}{\mathsf{e}}}
\newcommand{\south}[2]{\alldirections{#1}{#2}{\mathsf{s}}}
\newcommand{\north}[2]{\alldirections{#1}{#2}{\mathsf{n}}}
\newcommand{\sowe}[2]{\alldirections{#1}{#2}{\mathsf{sw}}}
\newcommand{\soea}[2]{\alldirections{#1}{#2}{\mathsf{se}}}
\newcommand{\nowe}[2]{\alldirections{#1}{#2}{\mathsf{nw}}}
\newcommand{\noea}[2]{\alldirections{#1}{#2}{\mathsf{ne}}}
\newcommand{\cent}[2]{\alldirections{#1}{#2}{\mathsf{c}}}
\newcommand{\wore}[2]{\alldirections{#1}{#2}{\mathsf{w,e}}}
\newcommand{\sorn}[2]{\alldirections{#1}{#2}{\mathsf{s,n}}}
\newcommand{\alphain}{\ensuremath{\alpha\in\{\mathsf{w,e,s,n}\}}}
\newcommand{\betain}{\ensuremath{\beta\in\{\mathsf{sw,se,nw,ne}\}}}

\newcommand{\fname}[1]{\mbox{\texttt{#1}}}                               
\newcommand{\formod}[1]{\mbox{\texttt{#1}}}                              
\newcommand{\forvar}[1]{\mbox{\texttt{#1}}}                              
\newcommand{\forsub}[1]{\mbox{\texttt{#1}}}                              

\begin{frontmatter}

\title{Two-Dimensional Central-Upwind Schemes for Curvilinear
       Grids and Application to Gas Dynamics with Angular Momentum}

\author[label1]{Tobias F. Illenseer\corauthref{cor}}
\ead{tillense@astrophysik.uni-kiel.de}
\corauth[cor]{Corresponding author.}
\author[label1,label2]{Wolfgang J. Duschl}
\ead{wjd@astrophysik.uni-kiel.de}

\address[label1]{Institut f\"ur Theoretische Physik und Astrophysik,
                 Leibnizstrasse 15, D-24118 Kiel, Christian-Albrechts-Universit\"at
                 zu Kiel, Germany}
\address[label2]{Steward Observatory, The University of Arizona, Tucson, Arizona 85721-0065}

\begin{abstract}
In this work we present new second order semi-discrete central schemes for
systems of hyperbolic conservation laws on curvilinear grids. Our methods
generalise the two-dimensional central-upwind schemes developed
by Kurganov and Tadmor \cite{kt:2002}. In these schemes we account for area and volume
changes in the numerical flux functions due to the non-cartesian geometries.
In case of vectorial conservation laws we introduce a general prescription of
the geometrical source terms valid for various orthogonal curvilinear coordinate
systems. The methods are applied to the two-dimensional Euler equations of inviscid
gas dynamics with and without angular momentum transport. In the latter case we
introduce a new test problem to examine the detailed conservation of specific
angular momentum.
\end{abstract}


\end{frontmatter}

\section{Introduction}
\label{sec:introduction}
In the last decades various multidimensional numerical schemes for the
solution of hyperbolic conservation laws have been developed. The continuous
increase in computing power provides the opportunity to perform accurate
computations even in the case of three-dimensional problems. However for
some physical systems this might not be the appropriate approach. In cases
where the underlying physics exhibit some kind of symmetry more accuracy can
be achieved by omitting the dependence on one dimension. In some cases these
symmetries might lead to different conservation laws. For instance in
rotationally symmetric inviscid fluids angular momentum is locally conserved.

Among the variety of numerical methods the family of Godunov-type schemes are
a very useful approach. Since the seminal work of Godunov \cite{go:1959}
these schemes became an important tool in the numerical analysis of hyperbolic
conservations laws. The original proposition of Godunov was quite simple:
Approximate the initial condition by piecewise constant data. Then advance the
solution in time by solving the intermittent local Riemann problems. Since those
days the method was improved by introducing less time-consuming approximate
Riemann-solvers. Furthermore the development of higher-order methods yields
better convergence of the numerical schemes. In most parts of our report we will
follow the work of Kurganov and Tadmor \cite{kt:2000,kt:2002}. They proposed
Riemann-solvers-free second order methods which avoid computation of the
eigensystem of the advection problem. We incorporate orthogonal curvilinear
coordinate systems into their cartesian scheme and therefore allow for area
and volume changes of the grid cells.

The outline of our paper is as follows: In Section~\ref{sec:conservation_laws}
we briefly review the covariant formulation of conservation laws and derive the
basic two-dimensional integral equation for general orthogonal coordinates. This
result is utilised in Section~\ref{sec:numerical_scheme} to obtain second order
semi-discrete non-oscillating numerical schemes resembling those described in
\cite{kt:2002}. In Section~\ref{sec:numerical_experiments} we present the results
of numerical computations obtained with the new scheme for the equations of gas
dynamics. The simulations were carried out on curvilinear meshes and on a
two-dimensional cartesian mesh for comparison. These tests examine the solution of
two-dimensional Riemann problems in polar and cylindrical symmetry. In addition
to that we introduce a method for testing the detailed conservation of angular
momentum for rotationally symmetric flows in Section~\ref{sec:angular_momentum}.
Finally a summary is given in Section~\ref{sec:conclusions}.

\subsection{Conservation laws and orthogonal curvilinear coordinate systems}
\label{sec:conservation_laws}

The concept of conservation is fundamental to a variety of physical phenomena
and leads to partial differential equations of almost the same kind in very
different areas. In this work we will focus on systems of non-linear
hyperbolic conservation laws of the form
\begin{equation}
  \del{u}{t} + \nabla\cdot T(u) = 0.
\end{equation}
Here $u$ is either a scalar or vector and $T$ a vector or rank-2 tensor.
$\nabla\cdot$ denotes the covariant derivative with respect to some
affine connection followed by a contraction over the last indices.
In the following we will restrict ourselves to orthogonal 
curvilinear coordinates $\{\xi,\eta,\phi\}$ with a local orthonormal
basis $\{\exi,\eeta,\ephi\}$ and metric scale factors $\{\hxi,\heta,\hphi\}$.
For the divergence of vector fields one obtains for all components with respect 
to local orthonormal frames
\begin{equation}
  \label{eqn:vector_divergence}
  \nabla\cdot v = \inv{\sqrt{g}} \biggl(\delxi{}\Bigl(\heta\hphi\vxi\Bigr)
    + \deleta{}\Bigl(\hxi\hphi\veta\Bigr) + \delphi{}\Bigl(\hxi\heta\vphi\Bigr)\biggr)
\end{equation}
and in case of tensor fields
\begin{equation}
  \label{eqn:tensor_divergence}
  \begin{aligned}
    \Bigl[\nabla\cdot T\Bigr]_{\hat{\xi}} &= \inv{\sqrt{g}}
      \biggl(\delxi{}\Bigl(\heta\hphi\Txx\Bigr)
        + \deleta{}\Bigl(\hxi\hphi\Txy\Bigr) + \delphi{}\Bigl(\hxi\heta\Txz\Bigr)
      \biggr) \\
      &\hspace{2.5cm} -\cyxy\Tyy - \czxz\Tzz + \cxyx\Tyx + \cxzx\Tzx \\
    \Bigl[\nabla\cdot T\Bigr]_{\hat{\eta}} &= \inv{\sqrt{g}}
      \biggl(\delxi{}\Bigl(\heta\hphi\Tyx\Bigr)
        + \deleta{}\Bigl(\hxi\hphi\Tyy\Bigr) + \delphi{}\Bigl(\hxi\heta\Tyz\Bigr)
      \biggr) \\
      &\hspace{2.5cm} -\cxyx\Txx - \czyz\Tzz + \cyxy\Txy + \cyzy\Tzy \\
    \Bigl[\nabla\cdot T\Bigr]_{\hat{\phi}} &= \inv{\sqrt{g}}
      \biggl(\delxi{}\Bigl(\heta\hphi\Tzx\Bigr)
        + \deleta{}\Bigl(\hxi\hphi\Tzy\Bigr) + \delphi{}\Bigl(\hxi\heta\Tzz\Bigr)
      \biggr) \\
      &\hspace{2.5cm} -\cxzx\Txx - \cyzy\Tyy + \czxz\Txz + \czyz\Tyz.
  \end{aligned}
\end{equation}
In these equations new geometrical quantities arise: The square root of the metric
determinant $\sqrt{g}=\hxi\heta\hphi$ and the commutator coefficients $c_{ijk}$ which depend
on the metric scale factors and their derivatives. A more detailed derivation is
given in the appendix.

In this paper we will focus on two-dimensional conservation laws by assuming
a coordinate symmetry with respect to $\phi$. We may consider the 3D case
in a follow-up paper. Hence we demand that all functions -- geometrical scale
factors as well as physical quantities -- are independent of $\phi$.
Therefore the commutator coefficients with $\hat{\phi}$ in their second index vanish
and a two-dimensional conservation law is obtained for the scalar
variable $u$ and vector field $v(u)$
\begin{equation}
  \label{eqn:cons_law2D_vector}
  \del{u}{t} + \inv{\sqrt{g}}
  \biggl(\delxi{}\Bigl(\heta\hphi\vxi(u)\Bigr)
  + \deleta{}\Bigl(\hxi\hphi\veta(u)\Bigr) \biggr) = 0.
\end{equation}
In the same way we derive a vectorial conservation law for the vector $w$ and
tensor field $T(w)$
\begin{equation}
  \label{eqn:cons_law2D_tensor}
  \begin{aligned}
    &\del{w_{\hat{\xi}}}{t} &+& \inv{\sqrt{g}}
      \biggl(\delxi{}\Bigl(\heta\hphi\Txx(w)\Bigr)
        + \deleta{}\Bigl(\hxi\hphi\Txy(w)\Bigr) \biggr) \\
    &&=& \cyxy\Tyy(w) + \czxz\Tzz(w) - \cxyx\Tyx(w) \\
    &\del{w_{\hat{\eta}}}{t} &+& \inv{\sqrt{g}}
      \biggl(\delxi{}\Bigl(\heta\hphi\Tyx(w)\Bigr)
        + \deleta{}\Bigl(\hxi\hphi\Tyy(w)\Bigr) \biggr) \\
    &&=& \cxyx\Txx(w) + \czyz\Tzz(w) - \cyxy\Txy(w) \\
    &\del{w_{\hat{\phi}}}{t} &+& \inv{\sqrt{g}}
      \biggl(\delxi{}\Bigl(\heta\hphi\Tzx(w)\Bigr)
        + \deleta{}\Bigl(\hxi\hphi\Tzy(w)\Bigr) \biggr) \\
    &&=& - \czxz\Txz(w) - \czyz\Tyz(w).
  \end{aligned}
\end{equation}
The only differences between scalar and vectorial conservation laws
are the geometrical source terms in case of the latter. Therefore
we can combine both to a system of conservation laws. At this point
it is convenient to define new spatial differential operators
\begin{equation}
\label{eqn:differential_operators}
\Dxi = \inv{\sqrt{g}}\delxi{}\heta\hphi,\qquad
\Deta = \inv{\sqrt{g}}\deleta{}\hxi\hphi
\end{equation}
and rewrite the conservation law 
\begin{equation}
  \label{eqn:cons_law2D_system}
  \partial_t u + \Dxi F(u) + \Deta G(u) = S(u).
\end{equation}
In this compact notation $u$ denotes the vector of conservative variables,
$F(u)$, $G(u)$ and $S(u)$ are the flux vectors and geometrical source terms,
respectively.

To allow for discontinuous solutions one integrates the differential
equation~(\ref{eqn:cons_law2D_system}) over the time interval
$[t_n,t_{n+1}]$ and a spatial region $D$ given by the
cartesian product $[\xi_-,\xi_+]\times[\eta_-,\eta_+]$.
Hence we obtain an integral equation, the so called \emph{weak formulation}
\begin{equation}
\label{eqn:integral_conservation}
\imean{u(t_{n+1})}_D = \imean{u(t_{n})}_D
- \int_{t_n}^{t_{n+1}}\imean{\Dxi F + \Deta G - S}_D\di{t}
\end{equation}
using the notation $\imean{~}$ for volume\footnote{The integration
with respect to $\phi$ is suppressed throughout the whole paper, because
all functions are considered independent of $\phi$. Nevertheless we will stick
to the term ``volume'' for integrals over two-dimensional domains.}
averages as defined in
\begin{equation}
  \label{eqn:volume_average}
  \imean{X(t)}_D =
    \inv{\DeltaV{}{}}\int_{\xi_-}^{\xi_+}
    \int_{\eta_-}^{\eta_+}
    X(t,\xi,\eta) \sqrt{g}\di{\xi}\di{\eta}
\end{equation}
with spatial volume $\DeltaV{}{}$ of region $D$
\begin{equation}
  \label{eqn:cell_volume}
  \DeltaV{}{} = \int_D\di{V}
  = \int_{\xi_-}^{\xi_+}
    \int_{\eta_-}^{\eta_+}\sqrt{g}\di{\xi}\di{\eta}.
\end{equation}
Equation~(\ref{eqn:integral_conservation}) describes the time evolution of
volume averaged conservative variables $u$ in region $D$. So
far it is impossible to evaluate the flux integrals without further knowledge
of the function $u(t,\xi,\eta)$ on the surface of $D$ and at time $t$
within the interval $[t_n,t_{n+1}]$. However in the next section we will derive
a numerical scheme which provides an approximation to these integrals.

\section{Numerical scheme}
\label{sec:numerical_scheme}

\subsection{Semi-discrete scheme for generalised orthogonal coordinates}
\label{sec:semidiscrete_scheme}

\begin{figure}[htb]
	\centering{\input{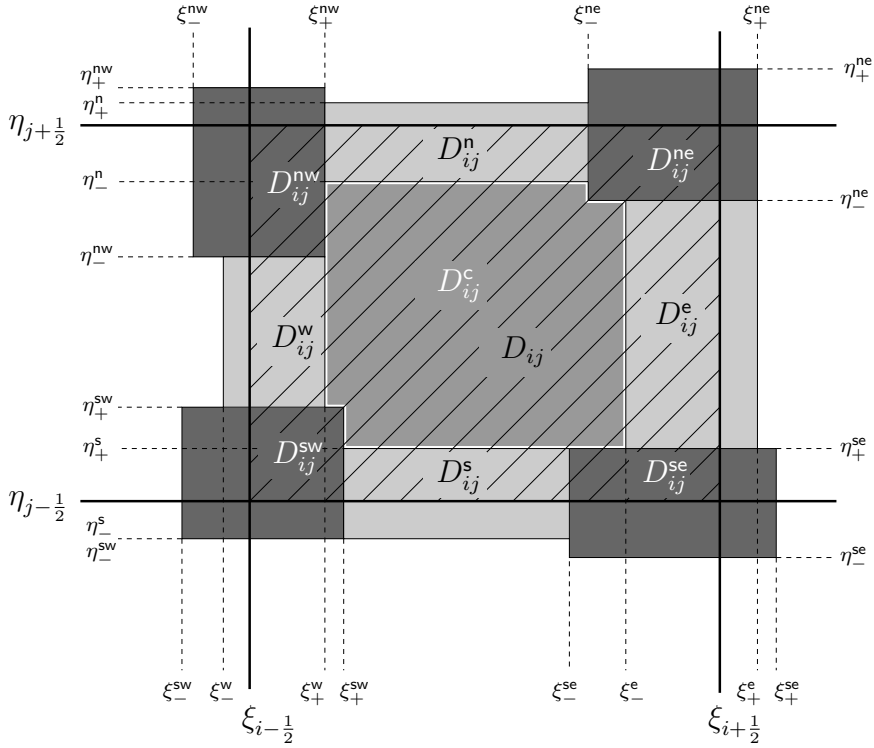}}
	\caption{Schematic view of the control volume}
	\label{fig:control_volume}
	\vspace*{1ex}
\end{figure}

The derivation of the numerical scheme follows the three steps of
reconstruction, evolution and projection described in \cite{kt:2002}.
For illustration consider the control volume selected by the cartesian product
$D_{i,j} = \bigl[\xi_{i-\thalf},\xi_{i+\thalf}\bigr]
\times\bigl[\eta_{j-\thalf},\eta_{j+\thalf}\bigr]$
in curvilinear orthogonal coordinates shown in Figure~\ref{fig:control_volume}.
There are two types of staggered grid cells drawn along the boundary:
\begin{description}
\item[Edge cells] are defined by the set union of partial regions at the edge
of two adjacent cells (light grey in Fig.~\ref{fig:control_volume}), e. ~g.\ at
the eastern boundary:
\begin{equation*}
D_{i+\thalf,j} = \east{D}{i,j}  \cup \west{D}{i+1,j}
\end{equation*}
whereas the partial areas addressed by the index pair $\{i,j\}$ are given by
\begin{align*}
\west{D}{i,j} &= \Bigl[\xi_{i-\thalf},\west{\xi}{+}\Bigr]
	\times\Bigl[\sowe{\eta}{+},\nowe{\eta}{-}\Bigr], &
\east{D}{i,j} &= \Bigl[\east{\xi}{-},\xi_{i+\thalf}\Bigr]
	\times\Bigl[\soea{\eta}{+},\soea{\eta}{-}\Bigr] \\
\south{D}{i,j} &= \Bigl[\sowe{\xi}{+},\soea{\xi}{-}\Bigr]
	\times\Bigl[\eta_{j-\thalf},\south{\eta}{+}\Bigr], &
\north{D}{i,j} &= \Bigl[\nowe{\xi}{+},\noea{\xi}{-}\Bigr]
	\times\Bigl[\north{\eta}{-},\eta_{j+\thalf}\Bigr].
\end{align*}
\item[Corner cells] are formed by the partial regions of the four
neighboring cells which meet in a cells corner (dark grey in
Fig.~\ref{fig:control_volume}), e.~g.\ around the 
south-eastern corner:
\begin{equation*}
D_{i+\thalf,j-\thalf}=\soea{D}{i,j} \cup \sowe{D}{i+1,j} 
	\cup \noea{D}{i,j-1} \cup \nowe{D}{i+1,j-1}.
\end{equation*}
With the help of the definition of all partial corner areas
with respect to cell $\{i,j\}$
\begin{align*}
\sowe{D}{i,j} &= \Bigl[\xi_{i-\thalf},\sowe{\xi}{+}\Bigr]
	\times\Bigl[\eta_{j-\thalf},\sowe{\eta}{+}\Bigr],&
\soea{D}{i,j} &= \Bigl[\soea{\xi}{-},\xi_{i+\thalf}\Bigr]
	\times\Bigl[\eta_{j-\thalf},\soea{\eta}{+}\Bigr] \\
\nowe{D}{i,j} &= \Bigl[\xi_{i-\thalf},\nowe{\xi}{+}\Bigr]
	\times\Bigl[\nowe{\eta}{-},\eta_{j+\thalf}\Bigr],&
\noea{D}{i,j} &= \Bigl[\noea{\xi}{-},\xi_{i+\thalf}\Bigr]
	\times\Bigl[\noea{\eta}{-},\eta_{j+\thalf}\Bigr].
\end{align*}
it is possible to construct the staggered corner cells.
\item[Central region:] Finally the remaining central region of
the cell is defined by the complement
\begin{equation}
\label{eqn:central_region}
\cent{D}{i,j} = \overline{\biggl(D_{i,j}\cap
	\bigcup_{\alpha\in\{\mathsf{w,e,s,n,sw,se,nw,ne}\}}
	D_{i,j}^{\alpha}\biggr)}.
\end{equation}
\end{description}
For the derivation of the numerical scheme we assume that at time
$t_n$ volume averages of the conservative variables
$u^n_{i,j}:=\imean{u(t_n)}_{D_{i,j}}$
are available for each cell in the computational domain.
Cell boundary data is obtained via piecewise linear reconstruction.
Using cell mean values and proper approximations for the slopes
the linear expansion yields
\begin{equation}
\label{eqn:linear_reconstruction}
\widetilde{u}_{i,j}^n(\xi,\eta) = u_{i,j}^n
	+ \bigl(\partial_{\xi}u\bigr)_{i,j}^n\bigl(\xi-\xi_0\bigr)
	+ \bigl(\partial_{\eta}u\bigr)_{i,j}^n\bigl(\eta-\eta_0\bigr),
\end{equation}
with $\xi,\xi_0\in\bigl[\xi_{i-\thalf},\xi_{i+\thalf}\bigr]$ and
$\eta,\eta_0\in\bigl[\eta_{j-\thalf},\eta_{j+\thalf}\bigr]$.
It is essential to deplete artificial oscillations caused by this
linear reconstruction process in order to obtain stability for
second order numerical schemes. Various methods are discussed in
the literature to achieve stable non-oscillating second order schemes,
see e.~g.\ \cite{vl:1977b,ha:1983,hll:1983,sw:1984}. The original
scheme by Kurganov and Tadmor follows the MUSCL \emph{(Monotone
Upstream-centred Schemes for Conservation Laws)} approach first
proposed by van Leer \cite{vl:1976}. This method introduces non-linear
functions -- so called (slope) limiters -- to damp spurious oscillations.
The method may also apply to a curvilinear scheme, if one demands consistency
with the averaging process
\begin{equation}
\label{eqn:reconstruction_constraint}
\imean{\widetilde{u}_{i,j}^n}_{D_{i,j}}
= \imean{u(t_n)}_{D_{i,j}} = u^n_{i,j}
\end{equation}
This equation should hold independently of the actual choice for the slopes.
M\"onchmeyer and M\"uller \cite{mm:1989} showed that this property is essential
to retain a conservative scheme in case of non-cartesian grids (see also
\cite{lh:1993}). With help of Eq.~(\ref{eqn:reconstruction_constraint}) one
derives the corollary that the coordinate pair $(\xi_0,\eta_0)$ is determined
by the barycentre of each control volume
\begin{equation}
\label{eqn:barycentre}
\xi_0 = \imean{\xi}_{D_{i,j}}\qquad\eta_0 = \imean{\eta}_{D_{i,j}}.
\end{equation}
Therefore cell mean values might be regarded as point values at the barycentre.
Equation~(\ref{eqn:linear_reconstruction}) together with an integral conservation
law for all the staggered grid cells similar to~(\ref{eqn:integral_conservation})
forms the building block to advance the solution in time. As an intermediate result
one computes cell mean values at time $t_{n+1}$ defined on the staggered grid. Finally
the updated mean values are obtained by reconstructing the staggered data and projecting
these functions onto the original cell area $D_{i,j}$
\begin{equation}
\label{eqn:projection}
u_{i,j}^{n+1} = \inv{\DeltaV{i,j}{}}\int_{D_{i,j}}
	\widetilde{w}^{n+1}(\xi,\eta)\di{V}
	= \imean{\widetilde{w}^{n+1}}_{D_{i,j}}
\end{equation}
This is the generalised formulation of equation (2.3) in \cite{kt:2002} for
orthogonal curvilinear coordinates. The function
$\widetilde{w}^{n+1}(\xi,\eta)$ refers to the combination of piecewise linear
reconstructions on the staggered grid. Depending on the control volume in which
they are defined this function may vary from cell to cell. Nevertheless
one can subdivide the integration into parts carried out over regions
determined by the intersection of $D_{i,j}$ with the staggered cells
(cf. Fig.~\ref{fig:control_volume})
\begin{equation}
\label{eqn:projection2}
\begin{split}
u_{i,j}^{n+1} &= \inv{\DeltaV{i,j}{}}
\Biggl(
	\sum_{\alphain}\DeltaV{i,j}{\alpha}
	\imean{\widetilde{w}_{\alpha}^{n+1}}_{D_{i,j}^{\alpha}}\\
     &+ \sum_{\betain}\DeltaV{i,j}{\beta}
	\imean{\widetilde{w}_{\beta}^{n+1}}_{D_{i,j}^{\beta}}
      + \DeltaV{i,j}{\mathsf{c}}
	\imean{\widetilde{w}_{\mathsf{c}}^{n+1}}_{\cent{D}{i,j}}
\Biggr).
\end{split}
\end{equation}
Up to this point no information about the advection problem has entered
our considerations. This changes if we fix the limits for integration
by introducing the minimal and maximal local speeds of propagation for
discontinuities according to \cite{knp:2001}, e.~g.\ at the western
and eastern cell boundaries
\newcommand{\lmin}{\lambda_{\mathsf{min}}}
\newcommand{\lmax}{\lambda_{\mathsf{max}}}
\begin{equation}
\begin{aligned}
\label{eqn:edge_wave_speeds}
a^+_{i\pm\thalf,j} = \max_{\omega\in\mathcal{C}
	\bigl(u_{i\pm\thalf,j}^+,u_{i\pm\thalf,j}^-\bigr)}
	\Biggl\{\lmax\biggl(\del{F}{u}(\omega)\biggr),0\Biggr\}\\
a^-_{i\pm\thalf,j} = \min_{\omega\in\mathcal{C}
	\bigl(u_{i\pm\thalf,j}^+,u_{i\pm\thalf,j}^-\bigr)}
	\Biggl\{\lmin\biggl(\del{F}{u}(\omega)\biggr),0\Biggr\}.
\end{aligned}
\end{equation}
Here $\mathcal{C}$ denotes a curve in phase space connecting two
adjacent states $u_{i-\thalf,j}^+$ and $u_{i-\thalf,j}^-$ of
neighboring cells via the Riemann fan and $\lmin,\lmax$ the
minimal and maximal eigenvalue of the Jacobian $\bigl(\del{F}{u}\bigr)$.
Similar definitions apply to the southern $b^{\pm}_{i,j-\thalf}$
and northern $b^{\pm}_{i,j+\thalf}$ wave speeds with the exception
that the Jacobian of $F$ has to be replaced by $\bigl(\del{G}{u}\bigr)$.
Although these expressions seem to be difficult to evaluate, for genuinely
nonlinear or linearly degenerate waves it is sufficient to compute
(cf.~\cite{knp:2001})
\begin{equation}
\begin{aligned}
\label{eqn:edge_wave_speeds1}
a^+_{i\pm\thalf,j} = \max\Biggl\{
	\lmax\biggl(\del{F}{u}\Bigl(u_{i\pm\thalf,j}^+\Bigr)\biggr),
	\lmax\biggl(\del{F}{u}\Bigl(u_{i\pm\thalf,j}^-\Bigr)\biggr),
	0\Biggr\}\\
a^-_{i\pm\thalf,j} = \min\Biggl\{
	\lmin\biggl(\del{F}{u}\Bigl(u_{i\pm\thalf,j}^+\Bigr)\biggr),
	\lmin\biggl(\del{F}{u}\Bigl(u_{i\pm\thalf,j}^-\Bigr)\biggr),
	0\Biggr\}.
\end{aligned}
\end{equation}
This is strictly true only for cartesian coordinates. The reason for this
limitation is that the underlying formalism is based on the characteristic
decomposition of the quasi-linear conservation law. In cartesian coordinates
the transformation to a quasi-linear form is straightforward. A simple calculation
with help of the chain rule yields
\begin{equation*}
\del{u}{t} + \biggl(\del{F}{u}\biggr)\del{u}{x}
	+ \biggl(\del{G}{u}\biggr)\del{u}{y} = 0.
\end{equation*}
However the curvilinear advection operators (\ref{eqn:differential_operators})
involve derivatives of geometrical scale factors and the quasi-linear
conservation law becomes
\begin{equation}
\label{eqn:cons_law2D_pseudo-linear}
\partial_t u + \biggl(\del{F}{u}\biggr)\Dxi u
	+ \biggl(\del{G}{u}\biggr)\Deta u = S(u).
\end{equation}
This is equivalent to Eq. (\ref{eqn:cons_law2D_system}) if and only if
the flux functions $F$ and $G$ are homogeneous functions of the conservative
variables $u$. For a homogeneous function
\begin{equation*}
F(u) = \biggl(\del{F}{u}\biggr) u
\end{equation*}
holds (cf. \cite{h1:1988} Chap.~16.2). Hence
\begin{equation*}
\begin{split}
	\Dxi F(u) &= \inv{\sqrt{g}}\delxi{}\Bigl(\heta\hphi F(u)\Bigr)
	= \inv{\sqrt{g}}\delxi{}\Bigl(\hxi\heta\Bigr) \biggl(\del{F}{u}\biggr) u
	+ \frac{\hxi\heta}{\sqrt{g}} \biggl(\del{F}{u}\biggr) \delxi{u} \\
	&=\biggl(\del{F}{u}\biggr) \inv{\sqrt{g}}\delxi{}\Bigl(\heta\hphi u\Bigr)
	= \biggl(\del{F}{u}\biggr) \Dxi u
\end{split}
\end{equation*}
and (\ref{eqn:cons_law2D_system}) may be rewritten in the quasi-linear form
(\ref{eqn:cons_law2D_pseudo-linear}). Apart from this discrepancy there is
another pitfall when using curvilinear coordinates. The extent of the staggered
grid cells is computed via multiplication of the local propagation speeds with
the time step $\Delta t$. However in curvilinear coordinates spatial distances
are not equal to coordinate distances. Therefore proper distances should be
obtained via evaluation of path integrals in compliance with
\begin{equation}
\label{eqn:proper_distance}
\begin{split}
	&\int_{\xi_{i\pm\thalf}}^{\wore{\xi}{\pm}} \hxi(\xi,\eta)\di{\xi}
	\approx \hxi(\xi_{i\pm\thalf},\eta)\bigl(\wore{\xi}{\pm}-\xi_{i\pm\thalf}\bigr)\\
	&\int_{\eta_{j\pm\thalf}}^{\sorn{\eta}{\pm}} \heta(\xi,\eta)\di{\eta}
	\approx \heta(\xi,\eta_{j\pm\thalf})\bigl(\sorn{\eta}{\pm}-\eta_{j\pm\thalf}\bigr).
\end{split}
\end{equation}
The accuracy of the approximations is sufficient as long as the coordinate
distance is small enough. In fact in the limit $\Delta t\to0$ the staggered
grid cells will collapse to lines, so that the considerations will hold.
Hence the limits of staggered zones along the edges with respect to
cell $D_{i,j}$ are given by
\begin{equation}
\label{eqn:face_limits}
\begin{aligned}
	\west{\xi}{-} &= \xi_{i-\thalf} 
	+ \frac{a^-_{i-\thalf,j}\Delta t}{\hxi(\xi_{i-\thalf},\eta)},&
	\west{\xi}{+} &= \xi_{i-\thalf} 
	+ \frac{a^+_{i-\thalf,j}\Delta t}{\hxi(\xi_{i-\thalf},\eta)},\\
	\east{\xi}{-} &= \xi_{i+\thalf}
	+ \frac{a^-_{i+\thalf,j}\Delta t}{\hxi(\xi_{i+\thalf},\eta)},&
	\east{\xi}{+} &= \xi_{i+\thalf}
	+ \frac{a^+_{i+\thalf,j}\Delta t}{\hxi(\xi_{i+\thalf},\eta)},\\
	\south{\eta}{-} &= \eta_{j-\thalf}
	+ \frac{b^-_{i,j-\thalf}\Delta t}{\heta(\xi,\eta_{j-\thalf})},&
	\south{\eta}{+} &= \eta_{j-\thalf}
	+ \frac{b^+_{i,j-\thalf}\Delta t}{\heta(\xi,\eta_{j-\thalf})},\\
	\north{\eta}{-} &= \eta_{j+\thalf}
	+ \frac{b^-_{i,j+\thalf}\Delta t}{\heta(\xi,\eta_{j+\thalf})},&
	\north{\eta}{+} &= \eta_{j+\thalf}
	+ \frac{b^+_{i,j+\thalf}\Delta t}{\heta(\xi,\eta_{j+\thalf})}.
\end{aligned}
\end{equation}
Whereas in the corners one computes
\begin{equation}
\label{eqn:corner_limits}
\begin{aligned}
	\sowe{\xi}{+} &= \xi_{i-\thalf}
	+ \frac{A^+_{i-\thalf,j-\thalf}\Delta t}{\hxi(\xi_{i-\thalf},\eta)},&
	\soea{\xi}{-}  &= \xi_{i+\thalf}
	+ \frac{A^-_{i+\thalf,j-\thalf}\Delta t}{\hxi(\xi_{i+\thalf},\eta)},\\
	\nowe{\xi}{+} &= \xi_{i-\thalf}
	+ \frac{A^+_{i-\thalf,j+\thalf}\Delta t}{\hxi(\xi_{i-\thalf},\eta)},&
	\noea{\xi}{-} &= \xi_{i+\thalf}
	+ \frac{A^-_{i+\thalf,j+\thalf}\Delta t}{\hxi(\xi_{i+\thalf},\eta)},\\
	\sowe{\eta}{+} &= \eta_{j-\thalf}
	+ \frac{B^+_{i-\thalf,j-\thalf}\Delta t}{\heta(\xi,\eta_{j-\thalf})},&
	\nowe{\eta}{-} &= \eta_{j+\thalf}
	+ \frac{B^-_{i-\thalf,j+\thalf}\Delta t}{\heta(\xi,\eta_{j+\thalf})},\\
	\soea{\eta}{+} &= \eta_{j-\thalf}
	+ \frac{B^+_{i-\thalf,j+\thalf}\Delta t}{\heta(\xi,\eta_{j-\thalf})},&
	\noea{\eta}{-} &= \eta_{j+\thalf}
	+ \frac{B^-_{i+\thalf,j+\thalf}\Delta t}{\heta(\xi,\eta_{j+\thalf})}.
\end{aligned}
\end{equation}
Here the propagation speeds are derived from the values of neighboring
cells according to (see \cite{knp:2001})
\begin{equation}
\label{eqn:corner_wave_speeds}
\begin{aligned}
	A^{+}_{i\pm\thalf,j-\thalf} &=
	\max\Bigl\{a_{i\pm\thalf,j}^+,a_{i\pm\thalf,j-1}^+\Bigr\},&
	A^{-}_{i\pm\thalf,j-\thalf} &=
	\min\Bigl\{a_{i\pm\thalf,j}^-,a_{i\pm\thalf,j-1}^-\Bigr\},\\
	A^{+}_{i\pm\thalf,j+\thalf} &=
	\max\Bigl\{a_{i\pm\thalf,j}^+,a_{i\pm\thalf,j+1}^+\Bigr\},&
	A^{-}_{i\pm\thalf,j+\thalf} &=
	\min\Bigl\{a_{i\pm\thalf,j}^-,a_{i\pm\thalf,j+1}^-\Bigr\},\\
	B^{+}_{i-\thalf,j\pm\thalf} &=
	\max\Bigl\{b_{i,j\pm\thalf}^+,b_{i-1,j\pm\thalf}^+\Bigr\},&
	B^{-}_{i-\thalf,j\pm\thalf} &=
	\min\Bigl\{b_{i,j\pm\thalf}^-,b_{i-1,j\pm\thalf}^-\Bigr\},\\
	B^{+}_{i+\thalf,j\pm\thalf} &=
	\max\Bigl\{b_{i,j\pm\thalf}^+,b_{i+1,j\pm\thalf}^+\Bigr\},&
	B^{-}_{i+\thalf,j\pm\thalf} &=
	\min\Bigl\{b_{i,j\pm\thalf}^-,b_{i+1,j\pm\thalf}^-\Bigr\}.
\end{aligned}
\end{equation}
With these definitions we may expand the integrals over staggered
cells arising in (\ref{eqn:projection2}), e. g. for the southern
domain $\south{D}{i,j}$ the integral of an arbitrary function $f(\xi,\eta)$
is given by
\begin{equation*}
\begin{aligned}
\int_{\south{D}{i,j}} f(\xi,\eta) \di{V}
&= \int_{\eta_{j-\thalf}}^{\south{\eta}{+}}
	\int_{\sowe{\xi}{+}}^{\soea{\xi}{-}}
	f(\xi,\eta)\hxi\heta\hphi\di{\xi}\di{\eta}\\
&= \int_{\sowe{\xi}{+}}^{\soea{\xi}{-}}
	f(\xi,\eta)\hxi\heta\hphi
	\bigl(\south{\eta}{+}-\eta_{j-\thalf}\bigr)
	\di{\xi}\bigg|_{\eta_{j-\thalf}}
	+ \order{\bigl(\south{\eta}{+}-\eta_{j-\thalf}\bigr)}{2}.
\end{aligned}
\end{equation*}
With the help of (\ref{eqn:face_limits}) this leads to
\begin{equation}
\label{eqn:southern_integral}
\int_{\south{D}{i,j}} f(\xi,\eta) \di{V}
= b_{i,j-\thalf}^{+}\Delta t
	\int_{\sowe{\xi}{+}}^{\soea{\xi}{-}}
	f(\xi,\eta)\hxi\hphi\di{\xi}\bigg|_{\eta_{j-\thalf}}
	+ \order{\Delta t}{2}.
\end{equation}
Therefore the volume of the southern region may be expressed by
\begin{equation}
\label{eqn:southern_volume}
\south{\Delta V}{i,j} = \int_{\south{D}{i,j}} \di{V}
= b_{i,j-\thalf}^{+}\Delta t
	\int_{\sowe{\xi}{+}}^{\soea{\xi}{-}}
	\hxi\hphi\di{\xi}\bigg|_{\eta_{j-\thalf}}
	+ \order{\Delta t}{2}.
\end{equation}
Furthermore it is necessary to compute approximations for the flux
integrals arising in (\ref{eqn:integral_conservation}), e.~g.\ again
for the southern domain this yields
\begin{align}
\label{eqn:southern_xflux_integral}
\int_{\south{D}{i,j}} \Dxi F \di{V}
&=	b_{i,j-\thalf}^{+}\Delta t
	\bigl[\hphi F\bigr]_{\sowe{\xi}{+},\eta_{j-\thalf}}^{\soea{\xi}{-},\eta_{j-\thalf}}
+	\order{\Delta t}{2}\\
\label{eqn:southern_yflux_integral}
\int_{\south{D}{i,j}} \Deta G \di{V}
&=	\Biggl[~
		\int_{\sowe{\xi}{+}}^{\soea{\xi}{-}}\hxi\hphi G\di{\xi}
	\Biggr]_{\eta_{j-\thalf}}^{\south{\eta}{+}}.
\end{align}
Similar equations can be obtained for the other staggered domains
along the edges. To proceed with the derivation of the numerical 
scheme we expand the first sum in (\ref{eqn:projection2}) up to
first order in $\Delta t$ using (\ref{eqn:southern_integral},
\ref{eqn:southern_volume})
\begin{equation}
\label{eqn:projection_first_sum}
\begin{split}
\sum_{\alphain}\DeltaV{i,j}{\alpha}
	\imean{\widetilde{w}_{\alpha}^{n+1}}_{D_{i,j}^{\alpha}}
&= \west{\Delta V}{i,j} w_{i-\thalf,j}^{n+1}
+ \east{\Delta V}{i,j} w_{i+\thalf,j}^{n+1}\\
&+ \south{\Delta V}{i,j} w_{i,j-\thalf}^{n+1}
+ \north{\Delta V}{i,j} w_{i,j+\thalf}^{n+1}
+ \order{\Delta t}{2}
\end{split}
\end{equation}
where $\widetilde{w}_{\alpha}^{n+1}$ denote the staggered reconstructions.
They are defined in the same way as the non-staggered reconstructions,
e.~g.\ in the southern domain according to
\begin{equation}
\label{eqn:staggered_reconstruction}
\begin{split}
\widetilde{w}_{\mathsf{s}}^{n+1}(\xi,\eta)
&= \widetilde{w}_{i,j-\thalf}^{n+1}(\xi,\eta)\Big|_{\xi,\eta\in\south{D}{i,j}}\\
&= w_{i,j-\thalf}^{n+1}
	+ \bigl(\partial_{\xi}w\bigr)_{i,j-\thalf}^{n+1}\bigl(\xi-\xi_0\bigr)
	+ \bigl(\partial_{\eta}w\bigr)_{i,j-\thalf}^{n+1}\bigl(\eta-\eta_0\bigr),
\end{split}
\end{equation}
with mean value $w_{i,j-\thalf}^{n+1}$ in the staggered area
$D_{i,j-\thalf}=\north{D}{i,j-1} \cup \south{D}{i,j}$. We would like to
emphasise that the cell barycentres $(\xi_0,\eta_0)$ in equation
\ref{eqn:staggered_reconstruction} are not identical to those in
equation \ref{eqn:barycentre} because they depend on the domain under
consideration. As in the case of non-staggered reconstructions we
demand consistency with the averaging process
[see~Eq.~(\ref{eqn:reconstruction_constraint})], i.~e.\
\begin{equation*}
\imean{\widetilde{w}_{i,j-\thalf}^{n+1}}_{D_{i,j-\thalf}}
= w_{i,j-\thalf}^{n+1}.
\end{equation*}
In case of the corner regions (second sum in \ref{eqn:projection2})
one can avoid the detailed computations. A short calculation proves
with help of Eq.~(\ref{eqn:corner_limits}) that all volume elements
are of order $\Delta t^2$ (cf.~\cite{knp:2001}), e.~g.\ for
the north-western domain $\nowe{D}{i,j}$
\begin{equation*}
  \begin{split}
    \nowe{\Delta V}{i,j} &= \int_{\nowe{D}{i,j}} \di{V}
  = \int_{\nowe{\eta}{-}}^{\eta_{j+\thalf}} \int_{\xi_{i-\thalf}}^{\nowe{\xi}{+}}
      \hxi\heta\hphi\di{\xi}\di{\eta} \\
  &= \Bigl(\eta_{j+\thalf}-\nowe{\eta}{-}\Bigr)
     \Bigl(\nowe{\xi}{+}-\xi_{i-\thalf}\Bigr)
     \hxi\heta\hphi\Big|_{\xi_{i-\thalf},\eta_{j+\thalf}} + \order{\Delta t}{4} \\
  &= -B_{i-\thalf,j+\thalf}^- A_{i-\thalf,j+\thalf}^+ \Delta t^2
     ~\hphi(\xi_{i-\thalf},\eta_{j+\thalf}) + \order{\Delta t}{4}.
  \end{split}
\end{equation*}
Therefore using piecewise lineare reconstructions similar to those in
(\ref{eqn:staggered_reconstruction}) one concludes that
\begin{equation}
\label{eqn:projection_second_sum}
\sum_{\betain}\DeltaV{i,j}{\beta}
	\imean{\widetilde{w}_{\beta}^{n+1}}_{D_{i,j}^{\beta}}
= \order{\Delta t}{2}.
\end{equation}
Henceforth one proceeds with the simplification of Eq.~(\ref{eqn:projection2}).
The expressions (\ref{eqn:projection_first_sum},\ref{eqn:projection_second_sum})
replace the first and second sum of (\ref{eqn:projection2}) and the updated cell
mean values become
\begin{equation}
\label{eqn:projection3}
\begin{split}
u_{i,j}^{n+1} = \inv{\DeltaV{i,j}{}}
\biggl(\west{\Delta V}{i,j} w_{i-\thalf,j}^{n+1}
&+ \east{\Delta V}{i,j} w_{i+\thalf,j}^{n+1}
+ \south{\Delta V}{i,j} w_{i,j-\thalf}^{n+1}\\
+ \north{\Delta V}{i,j} w_{i,j+\thalf}^{n+1}
&+ \cent{\Delta V}{i,j} w_{i,j}^{n+1}
\biggr)
+ \order{\Delta t}{2}
\end{split}
\end{equation}
Here we used the constraint (\ref{eqn:reconstruction_constraint})
to substitute $\imean{\widetilde{w}_{\mathsf{c}}^{n+1}}_{\cent{D}{i,j}}$
by the mean value $w_{i,j}^{n+1}$ within the central region.
The next step in the derivation of the update formula for cell mean values
incorporates the integral form of the conservation law. With help of
(\ref{eqn:integral_conservation}) one replaces the mean values on the
staggered grid at time step $t_{n+1}$ with flux and source term integrals.
The integrals with respect to time may then be approximated by the midpoint
quadrature rule. Hence for the central region one obtains
\begin{equation*}
\DeltaV{i,j}{\mathsf{c}}w_{i,j}^{n+1}
= \DeltaV{i,j}{\mathsf{c}}
\biggl(
	\imean{\widetilde{u}_{i,j}^n}_{\cent{D}{i,j}}
-	\imean{\Dxi F + \Deta G - S}_{\cent{D}{i,j}}\Big|_{t_n+\ttfrac{\Delta t}{2}}
	\Delta t + \order{\Delta t}{2}
\biggr).
\end{equation*}
To simplify integration over the irregular shaped domain $\cent{D}{i,j}$
one can substitute this integral with that over the whole cell area $D_{i,j}$
and subtract those integrals over the supplementary domains along the cell
boundary (see \ref{eqn:central_region})
\begin{equation*}
\begin{split}
\DeltaV{i,j}{\mathsf{c}}~w_{i,j}^{n+1}
	= \DeltaV{i,j}{}
	\biggl\{
		\imean{\widetilde{u}_{i,j}^n}_{D_{i,j}}
	&-	\imean{\Dxi F + \Deta G - S}_{D_{i,j}}
		\Big|_{t_n+\ttfrac{\Delta t}{2}}\Delta t
	\biggr\}\\
	- \sum_{\alphain}\DeltaV{i,j}{\alpha}
	\biggl\{
		\imean{\widetilde{u}_{i,j}^n}_{D_{i,j}^{\alpha}}
	&-	\imean{\Dxi F + \Deta G - S}_{D_{i,j}^{\alpha}}
			\Big|_{t_n+\ttfrac{\Delta t}{2}}\Delta t
	\biggr\}
	+ \order{\Delta t}{2}.
\end{split}
\end{equation*}
Again it is safe to omit the integrals over the corner areas (see
\ref{eqn:projection_second_sum}). Since some of the flux integrals over 
edge domains are of order $\Delta t$ (cf.~Eq.~\ref{eqn:southern_xflux_integral},
\ref{eqn:southern_yflux_integral}), multiplication with $\Delta t$ again leads
to terms of order $\Delta t^2$. Thus dropping all terms of order $\Delta t^2$ the
contribution due to the central region reduces to
\begin{equation*}
\begin{split}
\DeltaV{i,j}{\mathsf{c}}~w_{i,j}^{n+1}
	= \DeltaV{i,j}{}
	\biggl\{
		\imean{\widetilde{u}_{i,j}^n}_{D_{i,j}}
	&-	\imean{\Dxi F + \Deta G - S}_{D_{i,j}}
		\Big|_{t_n+\ttfrac{\Delta t}{2}}\Delta t
	\biggr\}\\
	- \sum_{\alphain}\DeltaV{i,j}{\alpha}
		\imean{\widetilde{u}_{i,j}^n}_{D_{i,j}^{\alpha}}
	&- \Delta t~
	\biggl\{
		\west{\Delta V}{i,j}\imean{\Dxi F}_{\west{D}{i,j}}
	+	\east{\Delta V}{i,j}\imean{\Dxi F}_{\east{D}{i,j}}\\
	+	\south{\Delta V}{i,j}\imean{\Deta G}_{\south{D}{i,j}}
	&+	\north{\Delta V}{i,j}\imean{\Deta G}_{\north{D}{i,j}}
	\biggr\}~\bigg|_{t_n+\ttfrac{\Delta t}{2}}
	+ \order{\Delta t}{2}.
\end{split}
\end{equation*}
This result is completely determined by reconstructed data inside
$D_{i,j}$ whereas for the staggered cells around the boundaries different
reconstructions of adjacent cells have to be taken into account.
The same argument regarding flux integrals over boundary areas as mentioned
above applies in this case. Thus the cell mean values within the edge regions
are determined by
\begin{align*}
\begin{split}
w_{i\pm\thalf,j}^{n+1}
=	&\inv{\DeltaV{i\pm\thalf,j}{}}
	\Biggl\{
		\DeltaV{i,j}{\mathsf{e(w)}}
		\biggl(
			\imean{\widetilde{u}_{i,j}^{n}}_{D_{i,j}^{\mathsf{e(w)}}}
		-	\imean{\Dxi F}_{D_{i,j}^{\mathsf{e(w)}}}
				\Big|_{t_n+\ttfrac{\Delta t}{2}}\Delta t
		\biggr)\\
	&+	\DeltaV{i\pm1,j}{\mathsf{w(e)}}
		\biggl(
			\imean{\widetilde{u}_{i\pm1,j}^{n}}_{D_{i\pm1,j}^{\mathsf{w(e)}}}
		-	\imean{\Dxi F}_{D_{i\pm1,j}^{\mathsf{w(e)}}}
				\Big|_{t_n+\ttfrac{\Delta t}{2}}\Delta t
		\biggr)
	\Biggr\}
+	\order{\Delta t}{2},\\
w_{i,j\pm\thalf}^{n+1}
=	&\inv{\DeltaV{i,j\pm\thalf}{}}
	\Biggl\{
		\DeltaV{i,j}{\mathsf{n(s)}}
		\biggl(
			\imean{\widetilde{u}_{i,j}^{n}}_{D_{i,j}^{\mathsf{n(s)}}}
		-	\imean{\Deta G}_{D_{i,j}^{\mathsf{n(s)}}}
				\Big|_{t_n+\ttfrac{\Delta t}{2}}\Delta t
		\biggr)\\
	&+	\DeltaV{i,j\pm1}{\mathsf{s(n)}}
		\biggl(
			\imean{\widetilde{u}_{i,j\pm1}^{n}}_{D_{i,j\pm1}^{\mathsf{n(s)}}}
		-	\imean{\Deta G}_{D_{i,j\pm1}^{\mathsf{n(s)}}}
				\Big|_{t_n+\ttfrac{\Delta t}{2}}\Delta t
		\biggr)
	\Biggr\}
+	\order{\Delta t}{2}.
\end{split}
\end{align*}
Insertion into (\ref{eqn:projection3}) yields
\begin{equation}
\label{eqn:projection4}
\begin{split}
u_{i,j}^{n+1} = u_{i,j}^n
-	\imean{\Dxi F + \Deta G - S}_{D_{i,j}} &
		\Big|_{t_n+\ttfrac{\Delta t}{2}}\Delta t\\
- \inv{\DeltaV{i,j}{}}
	\Biggl\{
		\frac{\west{\Delta V}{i,j}\east{\Delta V}{i-1,j}}{\DeltaV{i-\thalf,j}{}}
		\biggl(
			\imean{\widetilde{u}_{i,j}^n}_{\west{D}{i,j}}
		&-	\imean{\Dxi F}_{\west{D}{i,j}}\Big|_{t_n+\ttfrac{\Delta t}{2}}
			\Delta t\\
		-	\imean{\widetilde{u}_{i-1,j}^n}_{\east{D}{i-1,j}}
		&+	\imean{\Dxi F}_{\east{D}{i-1,j}}\Big|_{t_n+\ttfrac{\Delta t}{2}}
			\Delta t
		\biggr)\\
	+	\frac{\east{\Delta V}{i,j}\west{\Delta V}{i+1,j}}{\DeltaV{i+\thalf,j}{}}
		\biggl(
			\imean{\widetilde{u}_{i,j}^n}_{\east{D}{i,j}}
		&-	\imean{\Dxi F}_{\east{D}{i,j}}\Big|_{t_n+\ttfrac{\Delta t}{2}}
			\Delta t\\
		-	\imean{\widetilde{u}_{i+1,j}^n}_{\west{D}{i+1,j}}
		&+	\imean{\Dxi F}_{\west{D}{i+1,j}}\Big|_{t_n+\ttfrac{\Delta t}{2}}
			\Delta t
		\biggr)\\
	+	\frac{\south{\Delta V}{i,j}\north{\Delta V}{i,j-1}}{\DeltaV{i,j-\thalf}{}}
		\biggl(
			\imean{\widetilde{u}_{i,j}^n}_{\south{D}{i,j}}
		&-	\imean{\Deta G}_{\south{D}{i,j}}\Big|_{t_n+\ttfrac{\Delta t}{2}}
			\Delta t\\
		-	\imean{\widetilde{u}_{i,j-1}^n}_{\north{D}{i,j-1}}
		&+	\imean{\Deta G}_{\north{D}{i,j-1}}\Big|_{t_n+\ttfrac{\Delta t}{2}}
			\Delta t
		\biggr)\\
	+	\frac{\north{\Delta V}{i,j}\south{\Delta V}{i,j+1}}{\DeltaV{i,j+\thalf}{}}
		\biggl(
			\imean{\widetilde{u}_{i,j}^n}_{\north{D}{i,j}}
		&-	\imean{\Deta G}_{\north{D}{i,j}}\Big|_{t_n+\ttfrac{\Delta t}{2}}
			\Delta t\\
		-	\imean{\widetilde{u}_{i,j+1}^n}_{\south{D}{i,j+1}}
		&+	\imean{\Deta G}_{\south{D}{i,j+1}}\Big|_{t_n+\ttfrac{\Delta t}{2}}
			\Delta t
		\biggr)
	\Biggr\}
+ \order{\Delta t}{2}.
\end{split}
\end{equation}
Utilising (\ref{eqn:southern_integral}) and (\ref{eqn:southern_volume})
it is straightforward to prove that the contribution of all terms inside
the curly brackets is of order $\Delta t$. Therefore one can divide the
whole equation by $\Delta t$ and compute the limit $\Delta t\to 0$.
For convenience we define the time-dependent numerical fluxes across the
cell boundaries
\begin{equation}
\label{eqn:numerical_xfluxes}
\begin{split}
\mathcal{F}_{i+\thalf} = \inv{a_{i+\thalf,j}^+ - a_{i+\thalf,j}^-}~
	\int_{\eta_{j-\thalf}}^{\eta_{j+\thalf}}
	\biggl\{
		\Bigl(
			a_{i+\thalf,j}^+ F\bigl(\widetilde{u}_{i,j}\bigr)
		&-	a_{i+\thalf,j}^- F\bigl(\widetilde{u}_{i+1,j}\bigr)
		\Bigr)\\
	-	a_{i+\thalf,j}^+ a_{i+\thalf,j}^-
		\Bigl(
			\widetilde{u}_{i,j} &- \widetilde{u}_{i+1,j}  
		\Bigr)
	\biggr\} \heta\hphi\di{\eta}~\Bigg|_{\xi_{i+\thalf}}
\end{split}
\end{equation}
at the eastern edge and 
\begin{equation}
\label{eqn:numerical_yfluxes}
\begin{split}
\mathcal{G}_{j+\thalf} = \inv{b_{i,j+\thalf}^+ - b_{i,j+\thalf}^-}~
	\int_{\xi_{i-\thalf}}^{\xi_{i+\thalf}}
	\biggl\{
		\Bigl(
			b_{i,j+\thalf}^+ G\bigl(\widetilde{u}_{i,j}\bigr)
		&-	b_{i,j+\thalf}^- G\bigl(\widetilde{u}_{i,j+1}\bigr)
		\Bigr)\\
		-	b_{i,j+\thalf}^+ b_{i,j+\thalf}^-
		\Bigl(
			\widetilde{u}_{i,j} &- \widetilde{u}_{i,j+1} 
		\Bigr)
	\biggr\} \hxi\hphi\di{\xi}~\Bigg|_{\eta_{j+\thalf}}.
\end{split}
\end{equation}
at the northern edge respectively. Using 
(\ref{eqn:southern_integral}), (\ref{eqn:southern_volume}),
(\ref{eqn:southern_yflux_integral}) and similar equations for other boundaries
one derives the limits of all terms within the curly brackets in (\ref{eqn:projection4})
\begin{align*}
\pm\mathcal{F}_{i\pm\thalf}
	\mp	\int_{\eta_{j-\thalf}}^{\eta_{j+\thalf}}
		\heta\hphi F \di{\eta}~\Big|_{\xi_{i\pm\thalf}}
&=	\lim_{\Delta t\to 0}
	\frac{\DeltaV{i,j}{\mathsf{e(w)}}\DeltaV{i\pm1,j}{\mathsf{w(e)}}}
		{\DeltaV{i\pm\thalf,j}{}}
		\Biggl(
			\inv{\Delta t}\imean{\widetilde{u}_{i,j}^n}_{D_{i,j}^{\mathsf{e(w)}}}\\
		-	\imean{\Dxi F}_{D_{i,j}^{\mathsf{e(w)}}}\Big|_{t_n+\ttfrac{\Delta t}{2}}
		&-	\inv{\Delta t}\imean{\widetilde{u}_{i\pm1,j}^n}_{D_{i\pm1,j}^{\mathsf{w(e)}}}
		+	\imean{\Dxi F}_{D_{i\pm1,j}^{\mathsf{w(e)}}}\Big|_{t_n+\ttfrac{\Delta t}{2}}
		\Biggr),\\
\pm\mathcal{G}_{j\pm\thalf}
	\mp	\int_{\xi_{i-\thalf}}^{\xi_{i+\thalf}}
		\hxi\hphi G \di{\xi}~\Big|_{\eta_{j\pm\thalf}}
&=	\lim_{\Delta t\to 0}
	\frac{\DeltaV{i,j}{\mathsf{n(s)}}\DeltaV{i,j\pm1}{\mathsf{s(n)}}}{\DeltaV{i,j\pm\thalf}{}}
		\Biggl(
			\inv{\Delta t}\imean{\widetilde{u}_{i,j}^n}_{D_{i,j}^{\mathsf{n(s)}}}\\
		-	\imean{\Deta G}_{D_{i,j}^{\mathsf{n(s)}}}\Big|_{t_n+\ttfrac{\Delta t}{2}}
		&-	\inv{\Delta t}\imean{\widetilde{u}_{i,j\pm1}^n}_{D_{i,j\pm1}^{\mathsf{s(n)}}}
		+	\imean{\Deta G}_{D_{i,j\pm1}^{\mathsf{s(n)}}}\Big|_{t_n+\ttfrac{\Delta t}{2}}
		\Biggr).
\end{align*}
The remaining integrals on the left hand side of these equations cancel out with
the cell mean values of flux derivatives in (\ref{eqn:projection4}).
Finally we reach the semi-discrete update formula for the volume averaged
conservative variables within region $D_{i,j}$
\begin{equation}
\label{eqn:time_evolution}
\lim_{\Delta t\to 0}\frac{u_{i,j}^{n+1}-u_{i,j}^n}{\Delta t}
=	\dd{u_{i,j}}{t~}
=	-\frac{\mathcal{F}_{i+\thalf}-\mathcal{F}_{i-\thalf}}{\DeltaV{i,j}{}}
	-\frac{\mathcal{G}_{j+\thalf}-\mathcal{G}_{j-\thalf}}{\DeltaV{i,j}{}}
	+ \imean{S}_{D_{i,j}}
\end{equation}
with numerical flux functions given by (\ref{eqn:numerical_xfluxes})
and (\ref{eqn:numerical_yfluxes}). Cell mean values of source terms $S$ and
volume elements $\Delta V_{i,j}$ should be obtained by integration over the
domain $D_{i,j}$ according to (\ref{eqn:volume_average})
and~(\ref{eqn:cell_volume}).

Compared to the semi-discrete equation for time-evolution derived
in~\cite{kt:2002} our result offers a higher degree of generality in two ways:
\begin{enumerate}
\item The scheme applies to orthogonal coordinate systems and therefore
	accounts for area and volume changes of grid zones.
\item It utilises integral representations for the numerical fluxes
	instead of approximations with quadrature rules.
\end{enumerate}
The latter point allows us to discretise the numerical fluxes
and source terms in different ways starting from the same
integral formulas.

\subsection{Numerical approximations of flux and source term integrals}
\label{sec:numerical_fluxes}

The accuracy of the numerical scheme described in the previous section
is limited by the order of the polynomials in the reconstruction process.
Hence using quadrature rules of higher order to approximate the flux and
source term integrals do not improve the overall accuracy of the scheme.
Whereas quadrature rules of lower order would compromise the second order
accuracy of the scheme and lead to a larger numerical dissipation.
Therefore we will focus on second order quadrature schemes like the
midpoint and trapezoidal rule.

In the former case we define the reconstructed values at the cell interfaces
according to
\begin{equation}
\label{eqn:midpoint_reconstruction}
\begin{split}
	\begin{aligned}
		\west{u}{i,j}&=\widetilde{u}_{i,j}(\xi_{i-\thalf},\eta_j)&\qquad
		\south{u}{i,j}&=\widetilde{u}_{i,j}(\xi_i,\eta_{j-\thalf})\\
		\east{u}{i,j}&=\widetilde{u}_{i,j}(\xi_{i+\thalf},\eta_j)&\qquad
		\north{u}{i,j}&=\widetilde{u}_{i,j}(\xi_i,\eta_{j+\thalf})
	\end{aligned}
\end{split}
\end{equation}
and the numerical fluxes are given by
\begin{align}
\label{eqn:midpoint_xflux}
\mathcal{F}_{i+\thalf}^{\mathsf{mr}}
=	\frac{\Delta A_{i+\thalf,j}}{a_{i+\thalf,j}^+ - a_{i+\thalf,j}^-}
	\biggl\{
		a_{i+\thalf,j}^+ F(\east{u}{i,j})
	&-	a_{i+\thalf,j}^- F(\west{u}{i+1,j})\nonumber\\
	&-	a_{i+\thalf,j}^+ a_{i+\thalf,j}^-
		\bigl(
			\east{u}{i,j} - \west{u}{i+1,j}
		\bigr)
	\biggr\}\\
\label{eqn:midpoint_yflux}
\mathcal{G}_{j+\thalf}^{\mathsf{mr}}
=	\frac{\Delta A_{i,j+\thalf}}{b_{j+\thalf,j}^+ - b_{j+\thalf,j}^-}
	\biggl\{
		b_{j+\thalf,j}^+ G(\north{u}{i,j})
	&-	b_{j+\thalf,j}^- G(\south{u}{i,j+1})\nonumber\\
	&-	b_{j+\thalf,j}^+ b_{j+\thalf,j}^-
		\bigl(
			\north{u}{i,j} - \south{u}{i,j+1}
		\bigr)
	\biggr\}
\end{align}
with area elements being
\begin{equation}
\label{eqn:midpoint_areas}
\Delta A_{i+\thalf,j} = \heta\hphi\big|_{\xi_{i+\thalf},\eta_j}\Delta\eta,
\qquad
\Delta A_{i,j+\thalf} = \hxi\hphi\big|_{\xi_i,\eta_{j+\thalf}}\Delta\xi.
\end{equation}
The same rule applied to volume elements and source terms yields
\begin{equation}
\label{eqn:midpoint_volumes}
\Delta V_{i,j}^{\mathsf{mr}}
=	\hxi\heta\hphi\big|_{\xi_i,\eta_j}\Delta\xi\Delta\eta
\end{equation}
\begin{equation}
\label{eqn:midpoint_source}
\imean{S}_{D_{i,j}}^{\mathsf{mr}}
=	\inv{\DeltaV{i,j}{\mathsf{mr}}}~
	S(t,\xi_i,\eta_j)\hxi\heta\hphi\big|_{\xi_i,\eta_j}
	\Delta\xi\Delta\eta
=	S(t,\xi_i,\eta_j).
\end{equation}
In case of the trapezoidal rule we have to define the corner values as
\begin{equation}
\label{eqn:trapez_reconstruction}
\begin{split}
	\begin{aligned}
		\sowe{u}{i,j}&=\widetilde{u}_{i,j}(\xi_{i-\thalf},\eta_{j-\thalf})&\qquad
		\soea{u}{i,j}&=\widetilde{u}_{i,j}(\xi_{i+\thalf},\eta_{j-\thalf})\\
		\nowe{u}{i,j}&=\widetilde{u}_{i,j}(\xi_{i-\thalf},\eta_{j+\thalf})&\qquad
		\noea{u}{i,j}&=\widetilde{u}_{i,j}(\xi_{i+\thalf},\eta_{j+\thalf})
	\end{aligned}
\end{split}
\end{equation}
and obtain
\begin{align}
\label{eqn:trapez_xflux}
\mathcal{F}_{i+\thalf}^{\mathsf{tr}}
=	&\inv{2 \bigl(a_{i+\thalf,j}^+ - a_{i+\thalf,j}^-\bigr)}
	\Biggl\{
		\Delta A_{i+\thalf,j+\thalf}^{\hat{\xi}}
		\biggl(
			a_{i+\thalf,j}^+ F(\noea{u}{i,j})
		-	a_{i+\thalf,j}^- F(\nowe{u}{i+1,j})\nonumber\\
		&-	a_{i+\thalf,j}^+ a_{i+\thalf,j}^-
			\bigl(
				\noea{u}{i,j}-\nowe{u}{i+1,j}
			\bigr)
		\biggr)
	+	\Delta A_{i+\thalf,j-\thalf}^{\hat{\xi}}
		\biggl(
			a_{i+\thalf,j}^+ F(\soea{u}{i,j})\\
		&-	a_{i+\thalf,j}^- F(\sowe{u}{i+1,j})
		-	a_{i+\thalf,j}^+ a_{i+\thalf,j}^-
			\bigl(
				\soea{u}{i,j}-\sowe{u}{i+1,j}
			\bigr)
		\biggr)
	\Biggr\}\nonumber\\
\label{eqn:trapez_yflux}
\mathcal{G}_{j+\thalf}^{\mathsf{tr}}
=	&\inv{2\bigl(b_{i,j+\thalf}^+ - b_{i,j+\thalf}^-\bigr)}
	\Biggl\{
		\Delta A_{i+\thalf,j+\thalf}^{\hat{\eta}}
		\biggl(
			b_{i,j+\thalf}^+ G(\noea{u}{i,j})
		-	b_{i,j+\thalf}^- G(\soea{u}{i,j+1})\nonumber\\
		&-	b_{i,j+\thalf}^+ b_{i,j+\thalf}^-
			\bigl(
				\noea{u}{i,j}-\soea{u}{i,j+1}
			\bigr)
		\biggr)
	+	\Delta A_{i-\thalf,j+\thalf}^{\hat{\eta}}
		\biggl(
			b_{i,j+\thalf}^+ G(\nowe{u}{i,j})\\
		&-	b_{i,j+\thalf}^- G(\sowe{u}{i,j+1})
		-	b_{i,j+\thalf}^+ b_{i,j+\thalf}^-
			\bigl(
				\nowe{u}{i,j}-\sowe{u}{i,j+1}
			\bigr)
		\biggr)
	\Biggr\}.\nonumber
\end{align}
for the fluxes. Here the area elements are given by
\begin{align}
\label{eqn:trapez_xareas}
\Delta A_{i+\thalf,j+\thalf}^{\hat{\xi}}
&=	\heta\hphi\big|_{\xi_{i+\thalf},\eta_{j+\thalf}}\Delta\eta,\quad
\Delta A_{i+\thalf,j-\thalf}^{\hat{\xi}}
=	\heta\hphi\big|_{\xi_{i+\thalf},\eta_{j-\thalf}}\Delta\eta\\
\label{eqn:trapez_yareas}
\Delta A_{i+\thalf,j+\thalf}^{\hat{\eta}}
&=	\hxi\hphi\big|_{\xi_{i+\thalf},\eta_{j+\thalf}}\Delta\xi,\quad
\Delta A_{i-\thalf,j+\thalf}^{\hat{\eta}}
=\hxi\hphi\big|_{\xi_{i-\thalf},\eta_{j+\thalf}}\Delta\xi.
\end{align}
The volume elements yield
\begin{equation}
\label{eqn:trapez_volumes}
	\begin{split}
		\DeltaV{i,j}{\mathsf{tr}} = \frac{\Delta\xi\Delta\eta}{4}
		\biggl(
			&\hxi\heta\hphi\big|_{\xi_{i-\thalf},\eta_{j-\thalf}}
		+	\hxi\heta\hphi\big|_{\xi_{i+\thalf},\eta_{j-\thalf}}\\
		+	&\hxi\heta\hphi\big|_{\xi_{i-\thalf},\eta_{j+\thalf}}
		+	\hxi\heta\hphi\big|_{\xi_{i+\thalf},\eta_{j+\thalf}}
		\biggr).
	\end{split}
\end{equation}
and source terms, respectively
\begin{equation}
\label{eqn:trapez_source}
	\begin{split}
		\imean{S}_{D_{i,j}}^{\mathsf{mr}}
	= \frac{\Delta\xi\Delta\eta}{4~\DeltaV{i,j}{\mathsf{tr}}}
		\biggl(
			&S\,\hxi\heta\hphi\big|_{\xi_{i-\thalf},\eta_{j-\thalf}}
		+	S\,\hxi\heta\hphi\big|_{\xi_{i+\thalf},\eta_{j-\thalf}}\\
		+	&S\,\hxi\heta\hphi\big|_{\xi_{i-\thalf},\eta_{j+\thalf}}
		+	S\,\hxi\heta\hphi\big|_{\xi_{i+\thalf},\eta_{j+\thalf}}
		\biggr).
	\end{split}
\end{equation}
In the cartesian limit all scale factors are identical to one and
these formulas reduce to those derived in \cite{kt:2002}.

If the conservation law under consideration has a vectorial form,
inertial forces appear in the curvilinear description
(cf.~Eq.~\ref{eqn:cons_law2D_tensor}). To account for these geometrical
source terms we introduced the concept of commutator coefficients
in Section \ref{sec:conservation_laws}.
These non-linear functions depend on the scale factors
$\hxi,\heta,\hphi$ and their derivatives with respect to the
coordinates.
Numerical approximations of the commutator coefficients may in
principle involve Eqn.~(\ref{eqn:midpoint_source}) and (\ref{eqn:trapez_source}).
The remaining question is, how to approximate the derivatives of the
scale factors. A comparison of the truncation error of the source term integrals
with the flux differences shows that in case of the midpoint rule central differences
perform better whereas for the trapezoidal rule one-sided differences
lead to better results. Therefore the commutator coefficients for the midpoint rule
are given by
\begin{equation}
\label{eqn:midpoint_commutators}
\begin{split}
	\cyxy^{\mathsf{mr}} &= 
		\frac{\Delta\eta}{\DeltaV{i,j}{\mathsf{mr}}}
		\hphi(\xi_i,\eta_j)
		\Bigl(
			\heta\bigl(\xi_{i+\thalf},\eta_j\bigr)
		-	\heta\bigl(\xi_{i-\thalf},\eta_j\bigr)
		\Bigr) \\
	\cxyx^{\mathsf{mr}} &= 
		\frac{\Delta\xi}{\DeltaV{i,j}{\mathsf{mr}}}
		\hphi(\xi_i,\eta_j)
		\Bigl(
			\hxi\bigl(\xi_i,\eta_{j+\thalf}\bigr)
		-	\hxi\bigl(\xi_i,\eta_{j-\thalf}\bigr)
		\Bigr) \\
	\czxz^{\mathsf{mr}} &=
		\frac{\Delta\eta}{\DeltaV{i,j}{\mathsf{mr}}}
		\heta(\xi_i,\eta_j)
		\Bigl(
			\hphi\bigl(\xi_{i+\thalf},\eta_j\bigr)
		-	\hphi\bigl(\xi_{i-\thalf},\eta_j\bigr)
		\Bigr) \\
	\czyz^{\mathsf{mr}} &=
		\frac{\Delta\xi}{\DeltaV{i,j}{\mathsf{mr}}}
		\hxi(\xi_i,\eta_j)
		\Bigl(
			\hphi\bigl(\xi_i,\eta_{j+\thalf}\bigr)
		-	\hphi\bigl(\xi_i,\eta_{j-\thalf}\bigr)
		\Bigr).
\end{split}
\end{equation}
In case of the trapezoidal rule one requires four corner values for
each of the four commutator coefficients, i.e.
\begin{equation}
\label{eqn:trapez_commutators}
\begin{split}
	\cyxy^{\mathsf{tr,sw}} &=
		\inv{4}\frac{\Delta\eta}{\DeltaV{i,j}{\mathsf{tr}}}
		\hphi\bigl(\xi_{i-\thalf},\eta_{j-\thalf}\bigr)
		\Bigl(
			\heta\bigl(\xi_{i+\thalf},\eta_{j-\thalf}\bigr)
		-	\heta\bigl(\xi_{i-\thalf},\eta_{j-\thalf}\bigr)
		\Bigr) \\
	\cyxy^{\mathsf{tr,se}} &=
		\inv{4}\frac{\Delta\eta}{\DeltaV{i,j}{\mathsf{tr}}}
		\hphi\bigl(\xi_{i+\thalf},\eta_{j-\thalf}\bigr)
		\Bigl(
			\heta\bigl(\xi_{i+\thalf},\eta_{j-\thalf}\bigr)
		-	\heta\bigl(\xi_{i-\thalf},\eta_{j-\thalf}\bigr)
		\Bigr) \\
	\cyxy^{\mathsf{tr,nw}} &=
		\inv{4}\frac{\Delta\eta}{\DeltaV{i,j}{\mathsf{tr}}}
		\hphi\bigl(\xi_{i-\thalf},\eta_{j+\thalf}\bigr)
		\Bigl(
			\heta\bigl(\xi_{i+\thalf},\eta_{j+\thalf}\bigr)
		-	\heta\bigl(\xi_{i-\thalf},\eta_{j+\thalf}\bigr)
		\Bigr) \\
	\cyxy^{\mathsf{tr,se}} &=
		\inv{4}\frac{\Delta\eta}{\DeltaV{i,j}{\mathsf{tr}}}
		\hphi\bigl(\xi_{i+\thalf},\eta_{j+\thalf}\bigr)
		\Bigl(
			\heta\bigl(\xi_{i+\thalf},\eta_{j+\thalf}\bigr)
		-	\heta\bigl(\xi_{i-\thalf},\eta_{j+\thalf}\bigr)
		\Bigr) \\
\end{split}
\end{equation}
and similar expressions for the remaining commutator coefficients
$\cxyx, \czxz$ and $\czyz$. 

\section{Numerical Experiments}
\label{sec:numerical_experiments}

In the numerical examples we focus on the Euler equations for inviscid
compressible and non-heat-conducting gas dynamics. This system of non-linear
conservation laws may be written in generalised orthogonal coordinates according
to (\ref{eqn:cons_law2D_system}). Thereby we assumed symmetry of the flow with
respect to the spatial coordinate $\phi$. The vectors of conservative variables
and fluxes in the $\xi$ and $\eta$ directions are given by
\begin{equation}
\label{eqn:euler3d_fluxes}
u = \begin{bmatrix} \den \\ \den\vxi \\ \den\veta \\ \den\vphi \\ \etot \end{bmatrix},\quad
F = \begin{bmatrix} \den\vxi \\ \den\vxisq + \pre \\ \den\vxi\veta \\ \den\vxi\vphi \\
	(\etot+\pre)\vxi \end{bmatrix},\quad
G = \begin{bmatrix} \den\veta \\ \den\veta\vxi \\ \den\vetasq + \pre \\ \den\veta\vphi \\
	(\etot+\pre)\veta \end{bmatrix}.
\end{equation}
If we furthermore assume, that the equation of state is determined by the
ideal gas equation, the total energy $\etot$ depends on the density $\den$,
the pressure $\pre$ as well as the velocity components $\vxi$, $\veta$ and
$\vphi$, respectively
\begin{equation}
\label{eqn:total_energy}
\etot = \inv{2}\den\bigl(\vxisq+\vetasq+\vphisq\bigr) + \frac{\pre}{\gamma-1}.
\end{equation}
Here $\gamma$ denotes the constant ratio of specific heats. It is set
to $\gamma=1.4$ in all test configurations. For the geometrical source terms
one computes with the commutator coefficients
(\ref{eqn:comm_coeff})
\begin{equation}
\label{eqn:euler3d_sources}
S =	\den\vxi\begin{bmatrix}
		0 \\ -\veta\cxyx \\ \vxi\cxyx \\ -\vphi\czxz \\ 0
	\end{bmatrix}
  +	\den\veta\begin{bmatrix}
		0 \\ \veta\cyxy \\ -\vxi\cyxy \\ -\vphi\czyz \\ 0
	\end{bmatrix}
  +	\den\vphisq\begin{bmatrix} 0 \\ \czxz \\ \czyz \\ 0 \\ 0 \end{bmatrix}
  +	\pre\begin{bmatrix} 0 \\ \cyxy + \czxz \\ \cxyx + \czyz \\ 0 \\ 0 \end{bmatrix}.
\end{equation}
Besides these geometrical sources we do not account for additional body forces.
Therefore all units can be removed from the system given above. In all tests we
use this dimensionless prescription of the equations of gas dynamics.
In Sections \ref{sec:riemann2d_polar} and \ref{sec:riemann2d_cylindrical} we examine
Riemann problems with the initial data of $\vphi$ set to zero. In this case $\vphi$
remains zero and the fourth component of the vectors in (\ref{eqn:euler3d_fluxes})
and (\ref{eqn:euler3d_sources}) vanishes. The system of Euler equations
reduces to a pure two-dimensional problem without any flow in the direction
of $\phi$. In Section \ref{sec:angular_momentum} we study 
three-dimensional flows with rotational symmetry. The fourth equation of the
above mentioned system then describes the conservation of the angular momentum.

To advance the numerical solution in time according to (\ref{eqn:time_evolution})
we used a third order Runge-Kutta scheme as described in
\cite{sh:1988,so:1988}. Cell boundary data is obtained via piecewise linear
reconstruction. Unless stated otherwise we apply the \emph{monotonized-central}
limiter proposed in \cite{vl:1979}
\newcommand{\ldiff}{\ensuremath{\Delta_{i-\inv{2},j}}}
\newcommand{\rdiff}{\ensuremath{\Delta_{i+\inv{2},j}}}
\newcommand{\cdiff}{\ensuremath{\Delta_{i,j}}}
\begin{equation}
\label{eqn:mc_limiter}
\mc{\ldiff}{\rdiff}{\cdiff}
= \minmod{\theta\ldiff,\cdiff}{\theta\rdiff}
\end{equation}
with parameter $\theta\in[1,2]$. The left-handed, right-handed 
and central differences - e.~g. in the $\xi$-directions - are given by
\begin{equation*}
\ldiff= \frac{u_{i,j}-u_{i-\thalf,j}}{\Delta\xi},
\quad
\rdiff = \frac{u_{i+\thalf,j}-u_{i,j}}{\Delta\xi},
\quad
\cdiff=\frac{u_{i+\half,j}-u_{i-\thalf,j}}{2\Delta\xi},
\end{equation*}
and the multivariable \emph{minmod} function is evaluated according to
\begin{equation}
\label{eqn:minmod_limiter}
\minmod{x_1}{\ldots,x_n}
=	\begin{cases}
		\min(x_1,\ldots,x_n) & \textrm{if}~x_i>0~\forall\,i, \\
		\max(x_1,\ldots,x_n) & \textrm{if}~x_i<0~\forall\,i, \\
		0 & \textrm{otherwise}.
	\end{cases}
\end{equation}
The amount of numerical diffusion is controlled by the parameter
$\theta$. Lower values are more diffusive while higher
values result in an unstable scheme. In most of the test problems
we set the parameter $\theta=1.3$. This is less diffusive than
the ordinary minmod limiter (which corresponds to $\theta=1.0$) but
retains stability in most cases.

\subsection{Two-dimensional Riemann problem on a polar mesh}
\label{sec:riemann2d_polar}

In this section we analyse the numerical solutions of two-dimensional
Riemann problems on a polar mesh. The same tests were used by Kurganov
and Tadmor \cite{kt:2002} for comparison with the numerical results of
Schulz-Rinne et al.\ \cite{scg:1993} as well as Lax and Liu \cite{ll:1998}.
The computational domain is a square in the cartesian case and a circle for
the polar mesh with the origin in the centre in both cases.
At time $t=0$ the simulation is initialised with piecewise constant
data in the quadrants defined by the $x$ and $y$ axis.

Kurganov and Tadmor studied the numerical solutions of $19$ different configurations.
Since our numerical schemes differ only from those described in \cite{kt:2002}
by means of the geometrical factors, we will focus on the discrepancies due to the
polar mesh. Differences between computations on cartesian and polar grids would appear
in each of the $19$ test cases. Therefore we show only our results obtained for test
configuration number $18$ in \cite{kt:2002} with the initial data
\begin{align*}
  \pre_2 &= 1 &\den_2 &= 2    &\quad  \pre_1 &= 1 &\den_1 &= 1 \\
     u_2 &= 0 &   v_2 &= -0.3 &\quad     u_1 &= 0 &   v_1 &= 1 \\
  \\
  \pre_3 &= 0.4 &\den_3 &= 1.0625 &\quad  \pre_4 &= 0.4 &\den_4 &= 0.5197 \\
     u_3 &= 0   &   v_3 &= 0.2145 &\quad     u_4 &= 0   &   v_4 &= 0.2741.
\end{align*}
$u$ and $v$ denote the velocity in the $x$ and $y$ direction, respectively.
The north-eastern quadrant has index $1$, the others are labeled counterclockwise
in ascending order. These initial conditions generate a shock-wave between quadrants
2 and 3, a rarefaction-wave between quadrants 1 and 4 and contact discontinuities
between quadrants 1 and 2 as well as 3 and 4. In the centre of the computational domain
the 4 different solutions join leading to a complex flow.

The system of equations under consideration is given by (\ref{eqn:euler3d_fluxes}),
(\ref{eqn:total_energy}) and (\ref{eqn:euler3d_sources}) with $\vphi$ set to zero.
For cartesian coordinates all scale factors are unity and the commutator coefficients
and thus all geometrical source terms vanish. In case of the polar grid we identify
$\xi=r$, $\eta=\varphi$, $\phi=z$, respectively and assume slab symmetry, i.~e. all
derivatives with respect to $z$ are zero. The scale factors are $h_r=1$, $h_\varphi=r$
and $h_z=1$. The remaining non-vanishing commutator coefficient is therefore
\begin{equation*}
  \cyxy \equiv c_{\varphi r \varphi} = \inv{h_r h_\varphi}\del{h_\varphi}{r}=\inv{r}.
\end{equation*}
The computational domain covers a region of $1\times1$ in non-dimensional units with
a resolution of $400\times400$ cells on the cartesian mesh. In case of the polar
coordinates the extent of the circular domain is so large, that the unit square fits
exactly into it. Thus the radius of the computational domain is $\sqrt{2}/2$ with a
resolution of $282$ cells. The angular resolution is $360$ for the full $2\pi$ circular
domain. The parameter for the limiter was set to $\theta=1.3$ in all examples and the
Courant number is $0.4$. A stability criterion (CFL condition) similar to that proposed
by Courant, Friedrichs and Lewy \cite{cfl:1967} for an explicit central-upwind
scheme is given in \cite{kt:2000}.

\begin{figure}
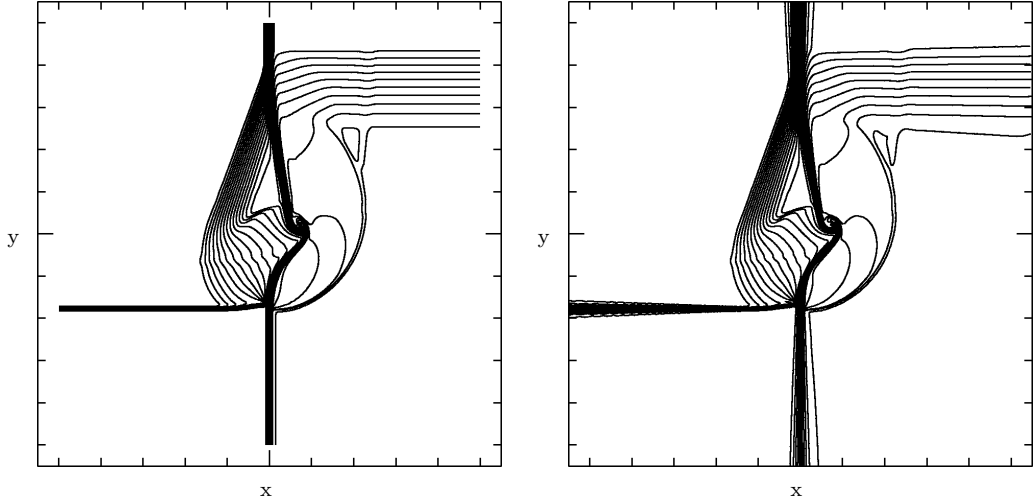

	\begin{minipage}[c]{0.5\textwidth}
		\centering
		\input{plots/riemann2d/KT-test18_400x400-midpnt-cons-mc1pt3.ptex}
	\end{minipage}
	\begin{minipage}[c]{0.5\textwidth}
		\centering
		\input{plots/riemann2d/KT-test18_282x360-midpnt-cons-mc1pt3-polar.ptex}
	\end{minipage}
	\caption{Density at time $t=0.2$ computed
		on a cartesian mesh (left) and polar mesh (right) using midpoint
		rule for flux calculations. $30$ equally spaced contour levels
		between $0.525$ and $2.025$.}
	\label{fig:KT-test18_midpnt}
	\hspace{1.0ex}
\end{figure}
\begin{figure}
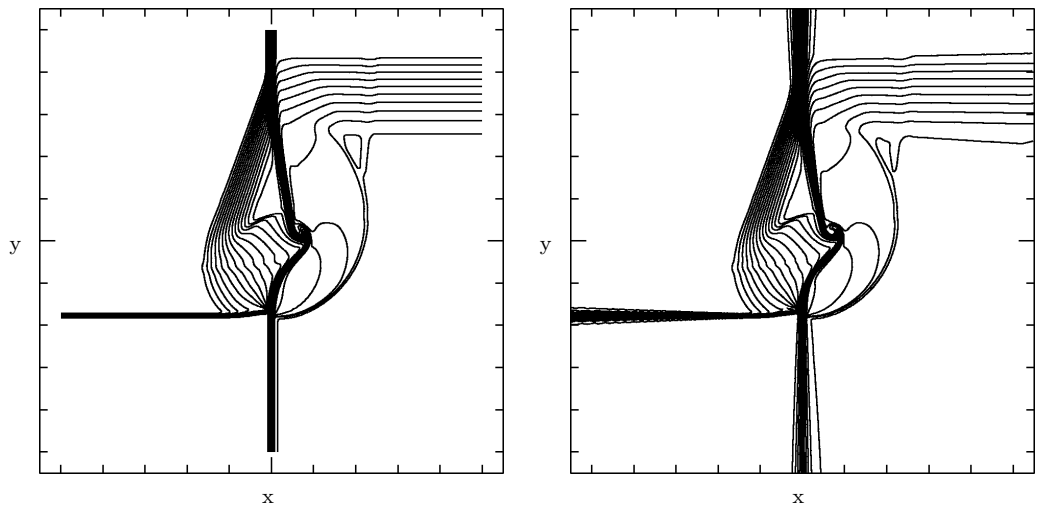

	\begin{minipage}[c]{0.5\textwidth}
		\centering
		\input{plots/riemann2d/KT-test18_400x400-trapez-cons-mc1pt3.ptex}
	\end{minipage}
	\begin{minipage}[c]{0.5\textwidth}
		\centering
		\input{plots/riemann2d/KT-test18_282x360-trapez-cons-mc1pt3-polar.ptex}
	\end{minipage}
	\caption{Density at time $t=0.2$ computed
		on a cartesian mesh (left) and polar mesh (right) using trapezoidal
		rule for flux calculations. $30$ equally spaced contour levels
		between $0.525$ and $2.025$.}
	\label{fig:KT-test18_trapez}
	\hspace{1.0ex}
\end{figure}

Figures \ref{fig:KT-test18_midpnt} and \ref{fig:KT-test18_trapez} show the density
contours for this two-dimensional Riemann problem. The numerical fluxes were calculated
utilising the midpoint rule (Fig.~\ref{fig:KT-test18_midpnt}) and the trapezoidal rule
(Fig.~\ref{fig:KT-test18_trapez}). There is almost no visible
difference in the dependency of the numerical fluxes. A quantitative analysis
shows relative deviations of the density distributions of less than $10^{-3}$ except
for the innermost area. Within a region of roughly $0.1$ around the centre one
observes a discrepancy between the solutions obtained with the two different
flux functions of a few percent. The results given in \cite{kt:2002}
for cartesian geometry are similar to ours besides the small 'bump' in the north-western
quadrant. In our results this 'bump' shows up only on small scales at a density level
of $\den=2.0125$.

In the simulations on the polar mesh we achieved results which are almost identical
to the cartesian case, at least in the central region. At a distance of $0.2$
from the centre the resolution of the shocks and contacts is less sharp on the
polar mesh and they tend to fan out towards the boundaries. The same applies to the
rarefactions in the north-eastern quadrant. The main reason for this may be
the decrease of the angular resolution with increasing distance to the centre.

\subsection{Spherical Riemann problem between walls on a cylindrical mesh}
\label{sec:riemann2d_cylindrical}

Langseth and LeVeque proposed a spherical Riemann problem in \cite{ll:2000}. The
flow under investigation  is rotationally symmetric and may be examined using
cylindrical coordinates. Therefore we identify $\xi=z$, $\eta=r$, $\phi=\varphi$
and assume symmetry with respect to the polar angle $\varphi$. The geometrical
scale factors are given by $h_z=1$, $h_r=1$ and $h_\varphi=r$. In this case the
only non-vanishing commutator coefficient is
\begin{equation*}
\czyz \equiv c_{\varphi r\varphi} = \inv{h_\varphi h_r}\del{h_\varphi}{r}
= \inv{r}.
\end{equation*}
The initial condition is a homogeneous density distribution $\den_0=1$ with
vanishing velocities. Inside a sphere with a radius of $r=0.2$ centred at $z=0.4$
the pressure is set to $5$ and outside to $1$. The fourth equation can be
removed from the system given by (\ref{eqn:euler3d_fluxes}), (\ref{eqn:total_energy})
and (\ref{eqn:euler3d_sources}), because initially there is no angular motion.
Hence $v_{\varphi}$ remains zero.
The computational domain covers a region of $[0,1.0]\times[0,1.5]$ in the
$r$-$z$-plane with a resolution of $400\times600$ cells. The parameter for the
\emph{monotonized-central} limiter is set to $\theta=1.3$ and the Courant number is
$0.4$ in all computations.

\begin{figure}
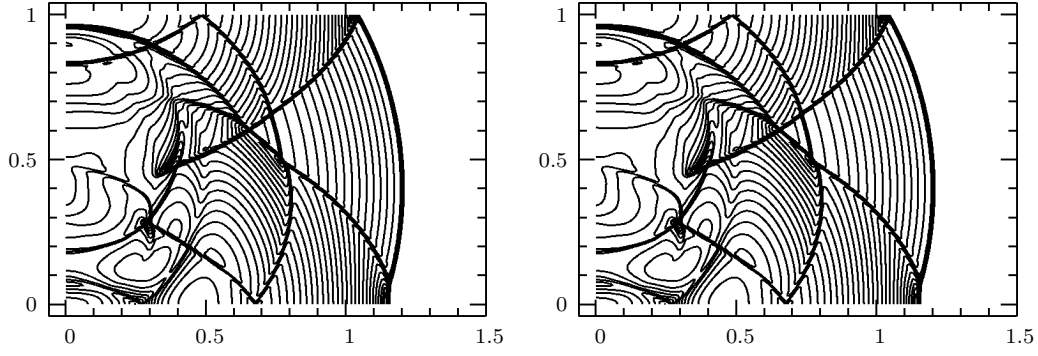

	\begin{minipage}[c]{0.5\textwidth}
		\centering
		\input{plots/leveque/leveque_400x600-midpnt-cons-mc1pt3-cyl.ptex}
	\end{minipage}
	\begin{minipage}[c]{0.5\textwidth}
		\centering
		\input{plots/leveque/leveque_400x600-trapez-cons-mc1pt3-cyl.ptex}
	\end{minipage}
    \caption{Pressure in the $r$-$z$ plane at $t=0.7$
		computed with different numerical fluxes using the midpoint rule (left) and
		the trapezoidal rule (right). $32$ equally spaced contour levels between
		$0.73$ and $1.48$.}
	\label{fig:leveque_pressure1}
	\hspace{1.0ex}
\end{figure}
\begin{figure}
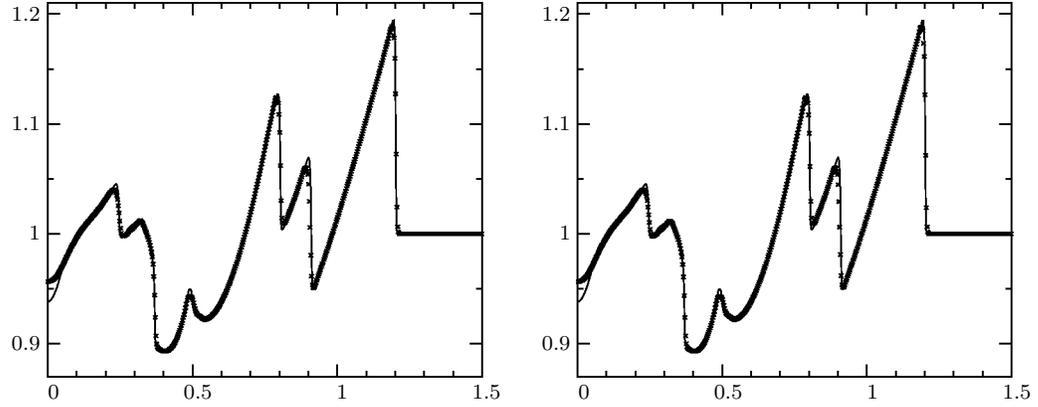

	\begin{minipage}[c]{0.5\textwidth}
		\centering
		\input{plots/leveque/leveque_400x600-cut160-midpnt-cons-mc1pt3-cyl.ptex}
	\end{minipage}
	\begin{minipage}[c]{0.5\textwidth}
		\centering
		\input{plots/leveque/leveque_400x600-cut160-trapez-cons-mc1pt3-cyl.ptex}
	\end{minipage}
    \caption{Horizontal profile of the pressure at $z=0.4$
		and time $t=0.7$ computed with different numerical fluxes using the midpoint rule
		(left) and the trapezoidal rule (right).}
	\label{fig:leveque_pressure2}
	\hspace{1.0ex}
\end{figure}
In Fig.~\ref{fig:leveque_pressure1} the pressure at time $t=0.7$ in the
$r$-$z$-plane is shown for the two different numerical fluxes. Both methods
seem to produce very similar results. The positions of the shocks are almost
identical to those computed in \cite{ll:2000}. There are small deviations
from their solutions in the central region along the axis. In this domain
the flow becomes stationary with low mass density. This discrepancy becomes more
obvious in the pressure profiles shown in Fig.~\ref{fig:leveque_pressure2}.
The solid lines in these diagrams correspond to a solution obtained on a finer
grid with a resolution of $800\times1200$ cells. In the limit $r\to 0$ the
pressure on the coarse grid is a little above $0.95$, but the solution converges
in both cases to the value stated in \cite{ll:2000} as the grid is refined.

Although there is almost no difference between the two methods regarding the
numerical results, we found that the scheme utilising the midpoint rule
is more robust in the high Mach number regime. In simulations with Mach
numbers above $100$ we observe that the trapezoidal scheme tends to steepen
pressure gradients which causes negative pressure in these remarkably high
supersonic flows.

\subsection{Detailed conservation of angular momentum}
\label{sec:angular_momentum}

The test introduced in this section makes use of an important property
of inviscid rotationally symmetric flows: The coupling of angular momentum
transport to mass transport. If one defines the
\emph{specific angular momentum} by
\begin{equation}
\label{eqn:spec_angular_momentum}
\sam = \hphi\vphi
\end{equation}
the fourth component of the system (\ref{eqn:cons_law2D_system}) with
(\ref{eqn:euler3d_fluxes}), (\ref{eqn:euler3d_sources}) may be rewritten
in the form
\begin{equation}
\label{eqn:angular_momentum_transport}
\partial_t\bigl(\den\sam\bigr) + \Dxi\bigl(\den\sam\vxi\bigr)
	+ \Deta\bigl(\den\sam\veta\bigr) = 0.
\end{equation}
This equation describes the advection of angular momentum density $\den\sam$ in
curvilinear orthogonal coordinates with rotational symmetry. It is of the very same
form as the equation for transport of mass density $\den$, i.~e.\ the continuity
equation
\begin{equation}
\label{eqn:continuity_equation}
\partial_t\den + \Dxi\bigl(\den\vxi\bigr) + \Deta\bigl(\den\veta\bigr) = 0.
\end{equation}
The time evolution of specific angular momentum is hence given by
\begin{equation}
\label{eqn:spec_angmom_transport}
\partial_t\sam = -\frac{\vxi}{\hxi}\,\partial_\xi \sam
	- \frac{\veta}{\heta}\,\partial_\eta \sam.
\end{equation}

\subsubsection{The mass spectrum as a constant of motion}
\label{sec:mass_spectrum}

Following the definition of Norman et al.\ \cite{nbw:1980} we define the mass spectrum
of the specific angular momentum by
\begin{equation}
\label{eqn:mass_spectrum}
M(\sam) = \int_0^\sam \di{m(\sam')}
\end{equation}
with $m(\sam')$ being the mass distribution function with respect
to specific angular momentum. For inviscid axisymmetric flows this
spectrum is a constant of motion within a spatial region $D$ if there
is no flux across its boundary $\partial D$
\begin{equation}
\label{eqn:massspec_constantofmotion}
\dd{M}{t}\Bigl|_D = 0 \qquad\textrm{if}\qquad
	\hat{n}\cdot v|_{\partial D}=0.
\end{equation}
$\hat{n}$ is the surface normal of the boundary and $v$ the velocity of
the fluid. With the help of the step function we may transform the
mass integral in (\ref{eqn:mass_spectrum}) into a volume integral
\begin{equation*}
\dd{M}{t}\Bigl|_D = \dd{}{t}\int_D \Theta\bigl(\sam-\sam'(\xi,\eta,t)\bigr)
	\,\den(\xi,\eta,t) \di{V}(\xi,\eta).
\end{equation*}
The spatial region $D$ does not depend on time, hence it can be exchanged
with the time derivative within the integration.
\begin{equation}
\label{eqn:proof_angmom_conservation1}
\dd{M}{t}\Bigl|_D
=	\int_D \partial_t\Theta(\sam-\sam')\,\den \di{V}
+	\int_D \Theta(\sam-\sam')\,\partial_t\den \di{V}
\end{equation}
In doing so we implied that the derivative of the step function is well defined
(see \cite{gs:1964}).
Equations (\ref{eqn:continuity_equation}) and (\ref{eqn:spec_angmom_transport})
allow us to replace the time derivatives with spatial ones. For convenience
we define the two-dimensional restrictions of the usual curvilinear differential
operators by
\newcommand{\nabhat}{\ensuremath{\widehat{\nabla}}}
\begin{equation*}
\nabhat\cdot v = \Dxi \vxi + \Deta \veta
\qquad\textrm{and}\qquad
v\cdot\nabhat \sam = \frac{\vxi}{\hxi}\,\partial_\xi \sam
	+ \frac{\veta}{\heta}\,\partial_\eta \sam.
\end{equation*}
Therefore (\ref{eqn:proof_angmom_conservation1}) becomes
\begin{align*}
\dd{M}{t}\Bigl|_D
&= 	- \int_D \den v\cdot\nabhat\sam'\,\dd{}{\sam'}\Theta(\sam-\sam')\di{V}
	- \int_D \Theta(\sam-\sam')\,\nabhat\cdot(\den v)\di{V}.\\
\intertext{The second integral may be evaluated by using integration
by parts and applying the Gaussian divergence theorem}
\dd{M}{t}\Bigl|_D
&=	- \int_D \den v\cdot\nabhat\Theta(\sam-\sam')\di{V}
	- \int_{\partial D} \den \Theta(\sam-\sam')\,v\cdot\hat{n}\di{A} \\
	&\quad+\int_D \den v\cdot\nabhat\Theta(\sam-\sam')\di{V}~ = ~0.
\end{align*}
The surface integral yields zero due to the vanishing normal velocity
across the boundary.

The mass spectrum $M(\sam)$ - although a global quantity - carries
information about the local redistribution of angular momentum. Any kind
of diffusive transport of the angular momentum causes a deviation from 
the initial spectrum. Since numerical diffusion exists in any numerical
scheme, we do not expect an exact conservation of the mass spectrum.

\subsubsection{Analytical expressions for the mass spectrum}
\label{sec:mass_spectrum_analytical}

In general the evaluation of the integral in (\ref{eqn:mass_spectrum}) is
difficult and must be done numerically. However, for a quantitative analysis
it is very useful for comparison to start numerical simulations with angular
momentum distributions for which analytical expressions of the mass spectrum
exist. Since cylindrical coordinates $\{z,r,\varphi\}$ exhibit the natural 
system for the formulation of rotationally symmetric problems, we will perform
all calculations in this system and assume symmetry with respect to $\varphi$.
Let us define the surface density by
\begin{equation}
\label{eqn:surface_density}
\Sden(r) = \int_{-\infty}^\infty \den(r,z)\di{z}
\end{equation}
and demand that this function is well defined for $r\in\mathbb{R}^+$.
If we claim that the specific angular momentum depends only on $r$, then
the mass spectrum (\ref{eqn:mass_spectrum}) may be written in terms of the
surface density
\begin{equation}
\label{eqn:mass_spectrum_simple1}
M(\sam) = 2\pi\int_0^\infty \Sden(r)\Theta\bigl((\sam-\sam'(r)\bigr)\, r \di{r}.
\end{equation}
Let us furthermore assume that $\sam'(r)$ is of the form
\begin{equation}
\label{eqn:specam_simple}
\sam'(r) = \sam_0 \biggl(\frac{r}{r_0}\biggr)^\alpha
\quad\textrm{with}\quad
\alpha,\sam_0,r_0 \in \mathbb{R}^+
\end{equation}
then the inverse function $r(\sam')=r_0\left(\sam'/\sam_0\right)^{1/\alpha}$ is well
defined and the mass spectrum becomes
\begin{equation}
\label{eqn:mass_spectrum_simple2}
M(\sam) = 2\pi\int_0^{r(\sam)} \Sden(r)\,r\di{r}.
\end{equation}
The last integral may be evaluated analytically in some cases. We will
consider mass distributions for which $\Sden(r)$ is a very simple
function of $r$:
\begin{enumerate}
\item homogeneous cylinder centred on the axis with radius $R$ and mass $M_0$
	\begin{equation}
	\label{eqn:mass_spectrum_cylinder}
		M(\sam) = M_0\,
			\biggl(\frac{r_0}{R}\biggr)^2
			\biggl(\frac{\sam}{\sam_0}\biggr)^{2/\alpha}
	\end{equation}
\item homogeneous sphere centred on the axis with radius $R$ and mass $M_0$
	\begin{equation}
	\label{eqn:mass_spectrum_sphere}
		M(\sam) = M_0\,
			\Biggl[1-\biggl(1-\biggl(\frac{r_0}{R}\biggr)^2
			\biggl(\frac{\sam}{\sam_0}\biggr)^{2/\alpha}\biggr)^{3/2}\Biggr] 
	\end{equation}
\item Gaussian density distribution centred on the axis with
standard deviation $\sigma$ and mass $M_0$ \\
	\begin{equation}
	\label{eqn:mass_spectrum_gauss}
		M(\sam) = M_0\,
		\Biggl[1-\exp{\biggl(
			-\inv{2}\biggl(\frac{r_0}{\sigma}\biggr)^2
			\biggl(\frac{\sam}{\sam_0}\biggr)^{2/\alpha}\biggr)}
		\Biggr].
	\end{equation}
\end{enumerate}
These expressions become even simpler for rigid motion. In this case
$\alpha=2$ and the specific angular momentum may be expressed in terms
of the constant angular velocity $\Omega_0$
\begin{equation*}
\sam(r) = \frac{\sam_0}{r_0^2}\,r^2 = \Omega_0\,r^2.
\end{equation*}
The spectrum of a rigidly rotating sphere (\ref{eqn:mass_spectrum_sphere})
then reduces to the formula given in \cite{nbw:1980}.

\subsubsection{The rigidly rotating Gaussian density distribution}
\label{sec:gaussian_pulse_cyl}

\begin{figure}
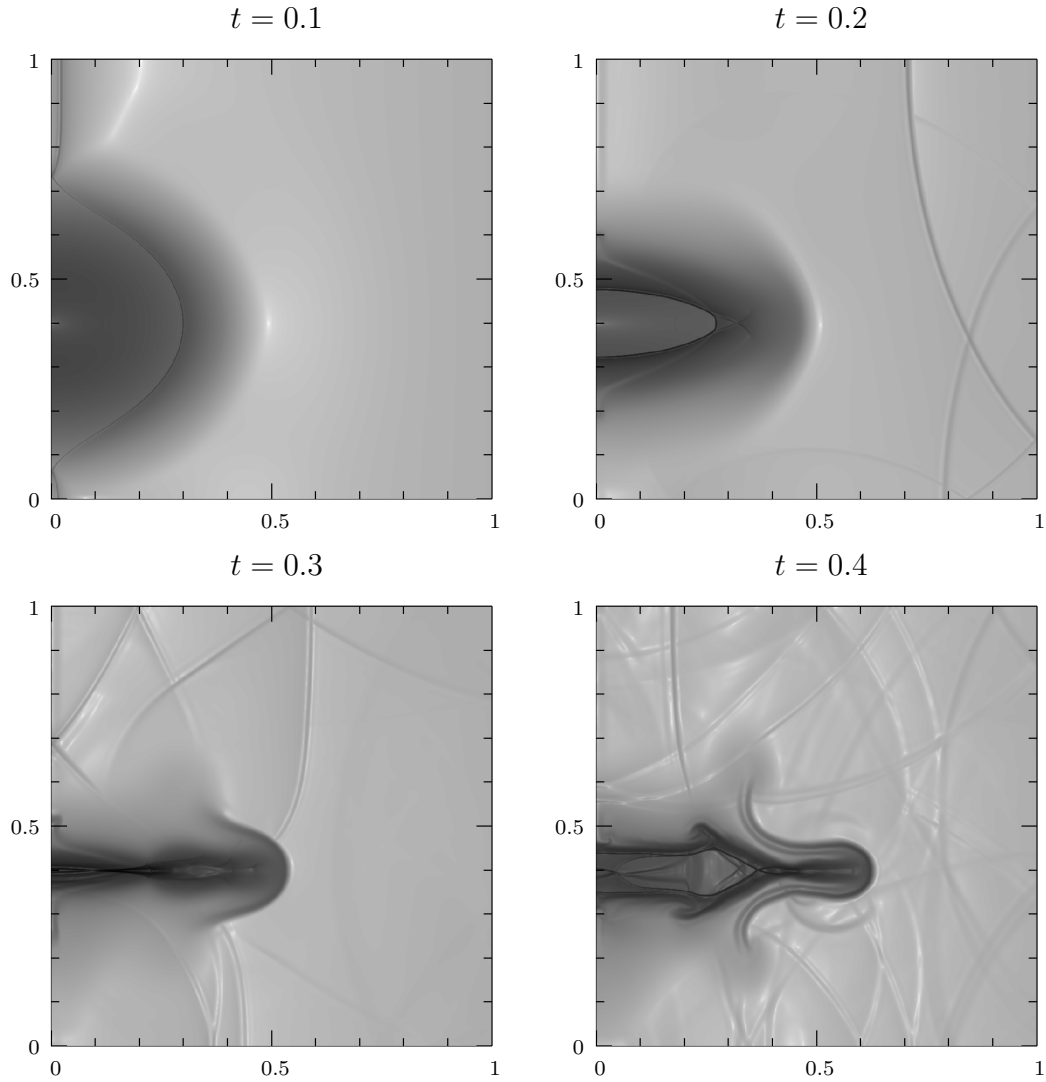

	\begin{minipage}[c]{0.48\textwidth}
		\input{plots/gauss3drot/graddensity-gauss3drot_800x800-midpnt-cons-mc1pt3-cyl_0001.ptex}
	\end{minipage}
	\hspace*{\fill}
	\begin{minipage}[c]{0.48\textwidth}
		\input{plots/gauss3drot/graddensity-gauss3drot_800x800-midpnt-cons-mc1pt3-cyl_0002.ptex}
	\end{minipage}
	
	\vspace*{2.0ex}
	\begin{minipage}[c]{0.48\textwidth}
		\input{plots/gauss3drot/graddensity-gauss3drot_800x800-midpnt-cons-mc1pt3-cyl_0003.ptex}
	\end{minipage}
	\hspace*{\fill}
	\begin{minipage}[c]{0.48\textwidth}
		\input{plots/gauss3drot/graddensity-gauss3drot_800x800-midpnt-cons-mc1pt3-cyl_0004.ptex}
	\end{minipage}
	\caption{Schlieren type images showing the absolute value of the density gradient
		at different times; logarithmic scale between $10^{-5}$ (white) and $10^4$ (black).
		 The solution was obtained on a $800\times800$ cylindrical grid.}
	\label{fig:gauss3drot_graddensity}
	\hspace{1.0ex}
\end{figure}
In this test we examine the disruption of a Gaussian density pulse
by centrifugal forces. A dense pulse is located on the axis of symmetry
within a low density environment. The rotational velocity in the whole
computational domain is initialised with constant angular velocity,
i~e.\ $v_\varphi = \Omega_0\,r^2$ with $\Omega_0 = 10$. All other velocity
components are set to zero and the pressure is unity. The peak density of
the pulse is $\den_\mathrm{max} = 10$ and the uniform density of the ambient
medium is $\den_\mathrm{min} = 10^{-2}$. The pulse is centred at $(r,z)=(0,0.4)$
with a full width half mean of $0.1$. On the cylindrical mesh the computational
domain covers a region of $[0,1]\times[0,1]$ with reflecting boundary conditions
at all boundaries. In addition we switched the sign of the rotational velocity
within the ghost zones along the axis of symmetry. All
computations in this chapter were performed using the midpoint quadrature scheme
with the monotonized central limiter and a Courant number of $0.4$. The limiters
parameter was set to $\theta=1.3$ unless stated otherwise.

\begin{figure}
	\begin{minipage}[b]{0.48\textwidth}
		\input{plots/gauss3drot/denspre-gauss3drot_800x800-cut320-midpnt-cons-mc1pt3-cyl_0001.ptex}
		\caption{Profile of mass density (solid line, left scale) and pressure (dashed line, right
			scale) at $z=0.4$ and time $t=0.1$.}
		\label{fig:gauss3drot_denspre_0001}
	\end{minipage}
 	\hspace*{\fill}
 	\begin{minipage}[b]{0.48\textwidth}
		\input{plots/gauss3drot/vrvphi-gauss3drot_800x800-cut320-midpnt-cons-mc1pt3-cyl_0001.ptex}
		\caption{Profile of radial velocity $v_r$ (solid line) and rotational
			velocity $v_\varphi$ (dashed line) at $z=0.4$ and time $t=0.1$.}
		\label{fig:gauss3drot_vrvphi_0001}
 	\end{minipage}
\end{figure}
\begin{figure}
	\begin{minipage}[b]{0.48\textwidth}
		\input{plots/gauss3drot/density-gauss3drot_800x800-midpnt-cons-mc1pt3-cyl_0004.ptex}
		\caption{Mass density and specific angular momentum (contours) at time $t=0.4$. The scale
			for the contour levels is logarithmic with basis $2$ starting at $2^{-9}$.}
		\label{fig:gauss3drot_density}
	\end{minipage}
	\hspace*{\fill}
	\begin{minipage}[b]{0.48\textwidth}
		\input{plots/gauss3drot/density-gauss3drot_800x800-cut320-midpnt-cons-mc1pt3-cyl_0004.ptex}
		\caption{Profile of mass density (solid line, left scale) and specific angular momentum
			(dashed line, right scale) at $z=0.4$ and time $t=0.4$.}
		\label{fig:gauss3drot_density-cut}
		\vspace*{2.7ex}
	\end{minipage}
\end{figure}
This problem setup ensures that a huge amount of mass with low specific angular
momentum spreads out into the whole computational domain. The dynamic behaviour
is, at least in the beginning, mostly driven by centrifugal forces. Hence the
dynamic time scale is dominated by the time of circulation, which is roughly
$1/\Omega_0 = 0.1$. Fig.~\ref{fig:gauss3drot_graddensity} depict the time evolution
of the flow up to $t=0.4$ for a resolution of $800\times800$ cells. The forces acting
on the density pulse accelerate the gas to supersonic speed in the radial direction
forming a bow shock. One clearly sees how the material of the dense region is driven
outwards. Within this central region behind the shock the peak density drops to $5.0$
at time $t=0.1$ and the pressure arrives at a constant value of $0.3787$
(cf.\ Fig.~\ref{fig:gauss3drot_denspre_0001}). Since the pressure gradient inside
this bubble confined by the shock is zero, there are no forces acting in the
vertical direction and $v_z$ remains zero. Furthermore the remarkable identity
$v_r=v_\varphi$ holds as can be seen for $z=0.4$ in Fig.~\ref{fig:gauss3drot_vrvphi_0001}.
At $t=0.2$ the density pulse has flattend to a disk like structure causing the low pressure
region to collapse roughly around $t=0.3$.

\begin{figure}
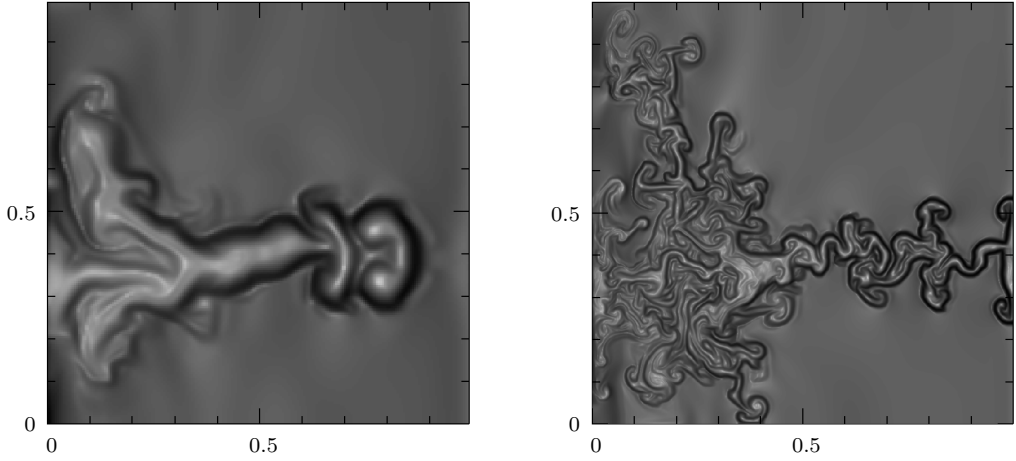

	\begin{minipage}[b]{0.48\textwidth}
		\input{plots/gauss3drot/vorticity-gauss3drot_200x200-midpnt-cons-mc1pt3-cyl_0010.ptex}
	\end{minipage}
 	\hspace*{\fill}
	\begin{minipage}[b]{0.48\textwidth}
		\input{plots/gauss3drot/vorticity-gauss3drot_800x800-midpnt-cons-mc1pt3-cyl_0010.ptex}
	\end{minipage}
	\caption{Absolute value of the projection of the vorticity onto the $r$-$z$-plane.
		Comparison of the results for two different resolutions; left image: $200\times200$,
		right image $800\times800$; both at time $t=1.0$.}
	\label{fig:vorticity-gauss3drot}
\end{figure}

Another interesting feature forms around $(r,z)=(0.5,0.4)$ which is very much like
the well known Rayleigh-Taylor instability \cite{ra:1883,ta:1950}. The dense material
of the pulse penetrates into the rarefied gas of the ambient medium, in this case driven
by centrifugal forces. These instabilities grow rapidly and the flow becomes
more and more turbulent from the time $t=0.4$ (q.~v.\ Fig.~\ref{fig:vorticity-gauss3drot}).

\begin{figure}
	\begin{minipage}[b]{0.48\textwidth}
		\input{plots/gauss3drot/amspec-gauss3drot_800x800-midpnt-cons-mc1pt3-cyl_0004.ptex}
		\caption{Mass spectrum of specific angular momentum for the initial condition and
			at time $t=0.4$; the resolution of the simulation is $800\times800$.}
		\label{fig:amspec-gauss3drot_800x800_0004}
	\end{minipage}
 	\hspace*{\fill}
	\begin{minipage}[b]{0.48\textwidth}
		\input{plots/gauss3drot/amspec-gauss3drot_midpnt-cons-mc1pt3-cyl_0010.ptex}
		\caption{Mass spectrum of specific angular momentum at time $t=1.0$ for
			different resolutions.}
		\vspace*{2.65ex}
		\label{fig:amspec-gauss3drot_0010}
	\end{minipage}
\end{figure}
The outwards driven material carries a vast amount of the mass initially concentrated
around $(r,z)=(0,0.4)$. Connected to this mass flux gas with low specific angular momentum
is transported to larger radii as can be seen in Fig.~\ref{fig:gauss3drot_density}
and \ref{fig:gauss3drot_density-cut}. The question under consideration is whether this
redistribution in space is solely advective or partly diffusive. Therefore
we compute the mass spectrum as described in the previous sections. In this test problem
the exact solution can be obtained via summation of the two solutions for the Gaussian density
pulse (Eq.~\ref{eqn:mass_spectrum_gauss}) and the homogeneous cylinder
(Eq.~\ref{eqn:mass_spectrum_cylinder}) with different total masses $M_0$. In
Fig.~\ref{fig:amspec-gauss3drot_800x800_0004} this analytical solution is depicted
in conjunction with the initial spectrum and the numerical result at time $t=0.4$. The steep
slope in this logarithmic diagram starting at $\sam=10^{-6}$ indicates the Gaussian pulse
whereas the kink at $\sam=10^{-1}$ marks the transition to the ambient medium. Besides some
spurious data points at the lower end of the spectrum, the numerical result is in good
agreement with the exact solution. Due to the limited resolution of the numerical computation,
there are some data points with different angular momentum attached to the same mass. This shows
up in both the initial setup and in the results for $t=0.4$. Also there is clearly a lower limit
for mass and specific angular momentum determind by the size of the grid cells and the distance
between the barycentres of the innermost cells and the axis.

\begin{figure}
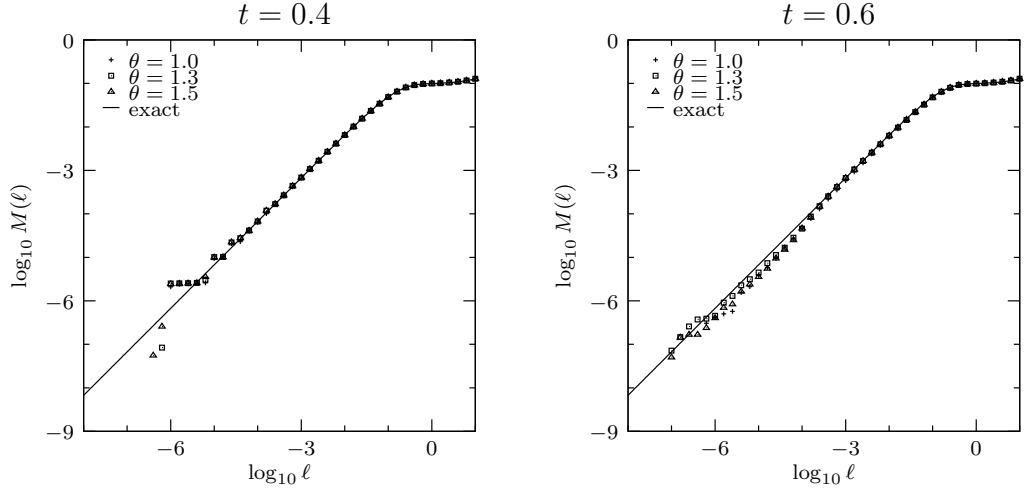

	\begin{minipage}[b]{0.48\textwidth}
		\input{plots/gauss3drot/amspec-gauss3drot_400x400-midpnt-cons-mc-cyl_0004.ptex}
	\end{minipage}
 	\hspace*{\fill}
	\begin{minipage}[b]{0.48\textwidth}
		\input{plots/gauss3drot/amspec-gauss3drot_400x400-midpnt-cons-mc-cyl_0006.ptex}
	\end{minipage}
	\caption{Mass spectrum of specific angular momentum. Comparison of the results
		for different settings of the monotonized-central limiter at different times $t$;
		the resolution of the simulations is $400\times400$.}
	\label{fig:amspec-gauss3drot_400x400}
\end{figure}
The picture becomes completely different, if we examine the data at later times. After
the generation of the first instabilities diffusive processes effect the redistribution
of angular momentum. This shows up in the loss of cells with low mass causing a steeper
slope in the spectrum seen in Fig.~\ref{fig:amspec-gauss3drot_0010}. Even more noteworthy
there is a dependence on the resolution, which seems to be contradictory. The simulations
with low resolution produce better results, whereas the numerical diffusion should
decrease with increasing resolution. To study the influence of the numerical diffusion
we changed the parameter of the monotonized-central limiter (see Eq.~\ref{eqn:mc_limiter}).
In general one observes that higher values of $\theta$ result in less diffusive
schemes. In Fig.~\ref{fig:amspec-gauss3drot_400x400} the mass spectrum before and after
the generation of the first instabilities is shown for different settings of the limiter.
Up to $t=0.4$ there is almost no dependence on the limiters parameter $\theta$. At later
times diffusive angular momentum transport occurs as in the previous experiment. However
a comparison of the mass spectra is questionable if instabilities are generated. In such
case a proposition regarding the conservation of angular momentum is of limited significance,
because the spatial redistribution of angular momentum proceeds differently.

To shed light on this phenomenon, we compared the vorticity of the numerical results at
the same time for different resolutions. In cylindrical coordinates with rotational
symmetry the components of the vorticity are given by
\begin{equation}
\label{eqn:vorticity_components}
w_z = -\del{v_\varphi}{z},\qquad
w_r = \inv{r}\del{}{r}\bigl(rv_\varphi\bigr)\qquad\textrm{and}\qquad
w_\varphi = \del{v_r}{z} - \del{v_z}{r}.
\end{equation}
With help of (\ref{eqn:spec_angmom_transport}) one easily verifies, that
\begin{equation}
\label{eqn:spec_angmom_vorticity}
\partial_t \sam = r\bigl(v_z w_r - v_r w_z\bigr)
\end{equation}
holds. Hence there is a close connection between the time evolution of
specific angular momentum and the projection of the vorticity onto the
$r$-$z$-plane. Fig.~\ref{fig:vorticity-gauss3drot} shows the absolute value
of the projected vorticity for simulations performed with different
resolutions. There is clearly much more turbulent motion in the high
resolution simulation. Lots of eddies have formed and the turbulence
seems to be much more evolved than in the results obtained for
lower resolution. Dark regions in this diagram are an indicator for
large vorticity and mark the surface where angular momentum is
exchanged mostly between grid cells. A comparison by eye shows that
this effective surface for angular momentum transport is much bigger
for high resolution simulations. Therefore diffusive transport of
angular momentum due to numerical diffusion becomes much more efficient.

\begin{figure}
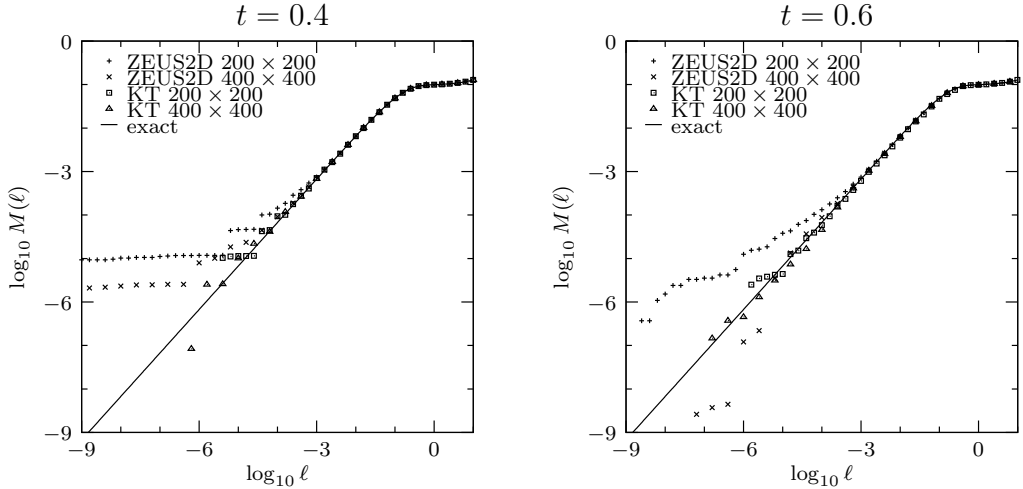

	\begin{minipage}[b]{0.48\textwidth}
		\input{plots/gauss3drot/amspec-gauss3drot_fosite-zeus2d-cyl_0004.ptex}
	\end{minipage}
 	\hspace*{\fill}
	\begin{minipage}[b]{0.48\textwidth}
		\input{plots/gauss3drot/amspec-gauss3drot_fosite-zeus2d-cyl_0006.ptex}
	\end{minipage}
	\caption{Mass spectrum of specific angular momentum. Comparison with the
		ZEUS2D code at different times $t$ for different resolutions.}
	\label{fig:amspec-gauss3drot_fosite-zeus2d}
\end{figure}
Finally we compare our numerical scheme to the well known second-order-accurate
van~Leer method \cite{vl:1977b} used in the ZEUS-2D code \cite{sn:1995} developed
by Stone and Norman \cite{sn:1992}. The ZEUS-2D code implements the numerical scheme
for consistent angular momentum transport proposed by Norman et al. \cite{nbw:1980}
and improved by Norman and Winkler \cite{nw:1986}. The results depicted in
Fig.~\ref{fig:amspec-gauss3drot_fosite-zeus2d} were computed with a von Neumann
and Richtmyer type scalar artificial viscosity \cite{vnr:1950}. The parameter which
controls the strength was set to $2.0$. For comparison we performed the same tests
using the tensorial artificial viscosity \cite{tw:1979}, but the results were almost
identical. The mass spectra obtained for these simulations show diffusive transport
in the low-mass and low-angular-momentum regime. We found that the ZEUS-2D code
generates grid zones with zero angular momentum indicated by the horizontal branch
in the left diagram of Fig.~\ref{fig:amspec-gauss3drot_fosite-zeus2d}. This leads
to an overestimation of the mass confined in grid zones with low specific angular
momentum. Henceforth the shape of the mass spectrum changes rapidly, when the first
instabilities appear, especially for higher resolution. By contrast, the mass spectrum 
computed for the new curvilinear central-upwind scheme does not show major deviations
from the exact solution for $t\leq0.6$.

\subsection{The Gaussian pulse on an oblate spheroidal mesh}
\label{sec:gaussian_pulse_obsph}

\begin{figure}
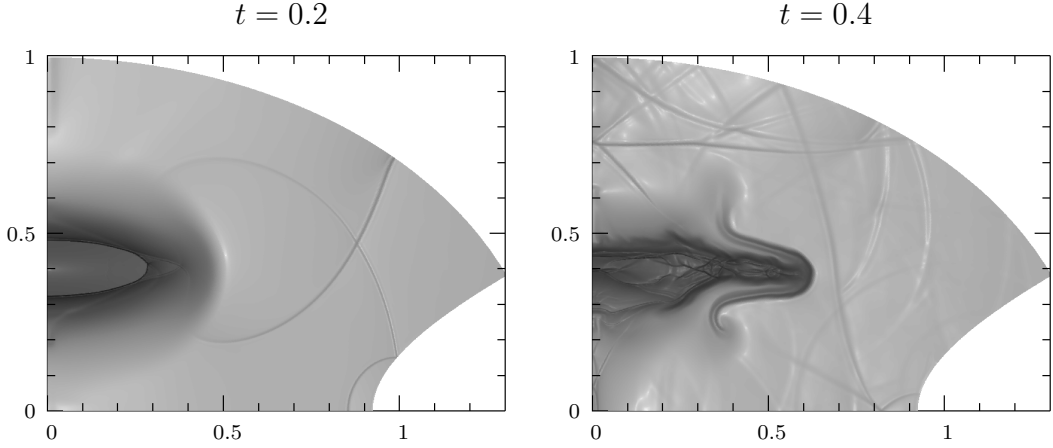

	\begin{minipage}[c]{0.48\textwidth}
		\input{plots/gauss3drot/graddensity-gauss3drot_800x800-midpnt-cons-mc1pt3-obsph_0002.ptex}
	\end{minipage}
	\hspace*{\fill}
	\begin{minipage}[c]{0.48\textwidth}
		\input{plots/gauss3drot/graddensity-gauss3drot_800x800-midpnt-cons-mc1pt3-obsph_0004.ptex}
	\end{minipage}
	\caption{Schlieren type images showing the absolute value of the density gradient 
		at two different times. The	the numerical solution was computed on an oblate
		spheroidal mesh with a resolution of $800\times800$ grid cells; scales are
		identical to those in Fig.~\ref{fig:gauss3drot_graddensity}.}
	\label{fig:gauss3drot_graddensity_obsph}
\end{figure}
The following test is meant as a demonstration for the validity of the
numerical scheme applied to other orthogonal and rotationally symmetric
coordinate systems. Hence we selected oblate spheroidal coordinates
 $\{\xi,\eta,\phi\}$ (see e.~g.~\cite{as:1964} chap.~21.3) with
\begin{equation}
\label{eqn:coordinates_obsph}
\begin{pmatrix} x \\ y \\ z \end{pmatrix}
= \begin{pmatrix} 
	a \tilde{\xi}\tilde{\eta}\cos{\phi} \\
	a \tilde{\xi}\tilde{\eta}\sin{\phi} \\
	a \sqrt{(\tilde{\xi}^2-1)(1-\tilde{\eta}^2)} \\
  \end{pmatrix}
= \begin{pmatrix} 
	a \cosh{\xi}\cos{\eta}\cos{\phi} \\
	a \cosh{\xi}\cos{\eta}\sin{\phi} \\
	a \sinh{\xi}\sin{\eta} \\
  \end{pmatrix}
\end{equation}
and metric scale factors 
\begin{equation}
\label{eqn:scale_factors_obsph}
\hxi=\heta=a\sqrt{(\sinh{\xi})^2+(\sin{\eta})^2},\qquad
\hphi=a\cosh{\xi}\cos{\eta}.
\end{equation}
The factor $a$ is a scaling constant which is set to unity in our calculations.
If we assume symmetry with respect to the coordinate $\phi$, the system of
two-dimensional Euler equations is again given by (\ref{eqn:cons_law2D_system})
with flux vectors (\ref{eqn:euler3d_fluxes}) and source terms
(\ref{eqn:euler3d_sources}). One easily proves by derivation of the scale factors
that none of the commutator coefficients vanishes. Therefore all geometrical
source terms have to be taken into account.

\begin{figure}
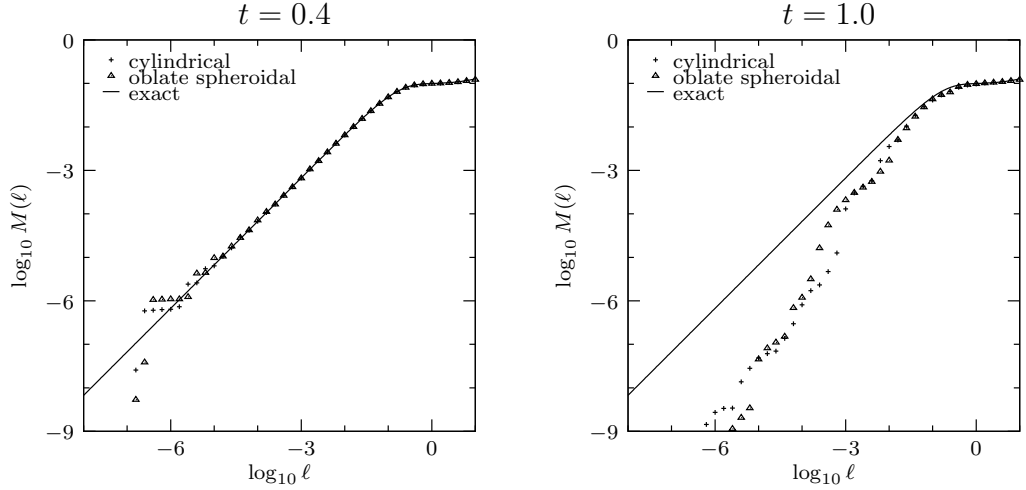

	\begin{minipage}[b]{0.48\textwidth}
		\input{plots/gauss3drot/amspec-gauss3drot_800x800-midpnt-cons-mc1pt3-obsph_0004.ptex}
	\end{minipage}
 	\hspace*{\fill}
	\begin{minipage}[b]{0.48\textwidth}
		\input{plots/gauss3drot/amspec-gauss3drot_800x800-midpnt-cons-mc1pt3-obsph_0010.ptex}
	\end{minipage}
	\caption{Mass spectrum of specific angular momentum. Comparison of the results
		for cylindrical and oblate spheroidal grids at different times $t$;
		the resolution of the simulations is $800\times800$.}
	\label{fig:amspec-gauss3drot_obsph}
\end{figure}
We examine the solution for the same initial condition as described in the previous
chapter. The Gaussian pulse with peak density $\den_\mathrm{max}=10$ is located on
the axis and embedded in an ambient medium with constant density
$\den_\mathrm{min}=10^{-2}$. The whole computational domain is initialised with constant
angular velocity $\Omega_0=10$ and constant pressure $\pre_0=1$. All boundaries are
treated as reflecting walls. The only difference to the test setup on the
cylindrical mesh is the shape of the computational domain with spatial
extension $(\xi,\eta)\in[0,0.88]\times[\pi/8,\pi/2]$. As for the cylindrical mesh the
resolution is $800\times800$.

For comparison with the simulations on cylindrical grids
(see Fig.~\ref{fig:gauss3drot_graddensity}) the numerical results for the density gradient
are depicted in Fig.~\ref{fig:gauss3drot_graddensity_obsph}. At time $t=0.2$ the solutions
on the oblate spheroidal mesh are almost identical to those on the cylindrical mesh. There
are small differences regarding the reflection of shock waves at the outer boundaries.
The influence of these disturbances on the main features becomes more important at later
times (see right diagram). Waves reflected backwards interact with the instabilities
mentioned in the previous chapter in a different way. Thus the solutions diverge
more and more. This does not affect the mass spectrum in a considerable manner
(see Fig~\ref{fig:amspec-gauss3drot_obsph}). As for the cylindrical grid we observe that
the spectrum is preserved very well up to $t=0.4$. Diffusive transport at later times
seems to alter the results in a similar way in either of the two curvilinear schemes.

\section{Conclusions}
\label{sec:conclusions}

The main advantage of central-upwind schemes is their simplicity. Apart from the
propagation speeds of the non-linear waves no other information about the
conservation laws under investigation is required. Therefore these schemes are
applicable to a variety of physical problems. Although the solution of the Euler
equations for gas dynamics is a major objective, one may also consider to solve the
equations for ideal magneto-hydrodynamics or the Hamilton-Jacobi equations.
(see \cite{btw:2004,bt:2006,knp:2001}). In this paper we extend this generality
to non-cartesian systems, thus providing the ability to solve systems of
hyperbolic conservation laws on orthogonal curvilinear grids.

A computer program written in \texttt{Fortran 95} has been developed in order to
test the new numerical schemes \cite{il:2007}. So far the program is capable of
solving the equations of inviscid gas dynamics in cartesian, polar, cylindrical,
spherical and oblate spheroidal coordinates. In case of rotational symmetry
the transport of angular momentum is included. Both quadrature rules for
computation of fluxes and source terms  are implemented. We performed
excessive tests to verify the correctness of the numerical results. Some of them
are presented in this paper. The solutions of two-dimensional Riemann problems
discussed in Sections~\ref{sec:riemann2d_polar} and~\ref{sec:riemann2d_cylindrical}
are in good agreement with the cartesian results presented by other authors
\cite{kt:2002,ll:2000,scg:1993}.

So far tests which check the detailed conservation of angular momentum are
very rare in the literature. Norman et al. \cite{nbw:1980} perform a test
similar to our setup, but they use a moving grid and include selfgravity
in their calculations. The significance of these results is limited due
to the fact that they study the mass spectrum at less than a time of circulation.
Therefore we proposed a pure hydrodynamical test with angular momentum transport
in Sec.~\ref{sec:gaussian_pulse_cyl}. A comparison with the ZEUS-2D code
\cite{sn:1995} demonstrates that our numerical scheme leads to better results
with less diffusive angular momentum transport.


\appendix
\renewcommand{\theequation}{A\arabic{equation}}
\section*{Appendix\quad Commutator coefficients}
\label{app:commutator_coefficients}

For orthogonal coordinate systems all off-diagonal elements in the metric
tensor vanish and the infinitesimal squared distance $\di{s}^2$ is
written in terms of metric coefficients and infinitesimal coordinate
displacements according to
\begin{equation}
  \label{eqn:inf_displacement1}
  \di{s}^2
  = g_{ij}\di{x^i}\di{x^j}
  = \bigl(\hxi\dxi\bigr)^2 + \bigl(\heta\deta\bigr)^2
  + \bigl(\hphi\dphi\bigr)^2.
\end{equation}
Hence one might define a set of orthonormal basis vectors by
\begin{equation}
  \exi = \inv{\hxi}\del{}{\xi} \qquad
  \eeta = \inv{\heta}\del{}{\eta} \qquad
  \ephi = \inv{\hphi}\del{}{\phi}
\end{equation}
and the corresponding set of dual 1-forms by
\begin{equation}
  \widetilde{\omega}^{\hat{\xi}} = \hxi\dxi \qquad
  \widetilde{\omega}^{\hat{\eta}} = \heta\deta \qquad
  \widetilde{\omega}^{\hat{\phi}} = \hphi\dphi.
\end{equation}
Furthermore one expands all vectors, tensors and forms with respect to the
new basis. Written in terms of orthonormal 1-forms the infinitesimal
squared distance becomes
\begin{equation}
  \label{eqn:inf_displacement2}
  \di{s}^2 = \bigl(\widetilde{\omega}^{\hat{\xi}}\bigr)^2
  + \bigl(\widetilde{\omega}^{\hat{\eta}}\bigr)^2 
  + \bigl(\widetilde{\omega}^{\hat{\phi}}\bigr)^2
  = \delta_{\hat{k}\hat{l}}~\widetilde{\omega}^{\hat{k}}\widetilde{\omega}^{\hat{l}}
\end{equation}
i.~e.\ the metric coefficients are independent of the coordinates. Therefore all
derivatives of the metric with respect to coordinates vanish and the
affine connection is determined by
\begin{equation}
  \label{eqn:connection}
  {\Gamma^{\hat{k}}}_{\hat{i}\hat{j}} = \inv{2} g^{\hat{k}\hat{l}}
  \bigl(c_{\hat{l}\hat{i}\hat{j}} + c_{\hat{l}\hat{j}\hat{i}} - c_{\hat{i}\hat{j}\hat{l}}\bigr)
\end{equation}
where $g^{\hat{k}\hat{l}}$ is the inverse metric and $c_{\hat{l}\hat{i}\hat{j}}$ denote the
commutator coefficients defined by means of the Lie-bracket
\begin{equation}
  \label{eqn:lie_bracket}
  \bigl[\widehat{e}_{\hat{i}},\widehat{e}_{\hat{j}}\bigr] 
  = {c_{\hat{i}\hat{j}}}^{\hat{k}}~\widehat{e}_{\hat{k}} 
  = g^{\hat{k}\hat{l}}~c_{\hat{i}\hat{j}\hat{l}}~\widehat{e}_{\hat{k}}.
\end{equation}
To obtain the commutator coefficients for the orthonormal basis mentioned 
above we have to apply the Lie-bracket of two basis vectors on an arbitrary
function and carry out the derivatives. 
\begin{equation*}
  \begin{split}
    \bigl[\exi,\eeta\bigr]~f(\xi,\eta,\phi)
      &= \inv{\hxi}\del{}{\xi}\biggl(\inv{\heta}\del{}{\eta}f\biggr)
      - \inv{\heta}\del{}{\eta}\biggl(\inv{\hxi}\del{}{\xi}f\biggr) \\
      &= \biggl(-\inv{\hxi\heta}\del{\heta}{\xi}\biggr)~\inv{\heta}\del{f}{\eta}
      + \biggl(\inv{\heta\hxi}\del{\hxi}{\eta}\biggr)~\inv{\hxi}\del{f}{\xi} \\
      &= \biggl(\cxyy~\eeta + \cxyx~\exi\biggr)~f
  \end{split}
\end{equation*}
The non-vanishing commutator coefficients are therefore given by
\begin{equation}
  \label{eqn:comm_coeff}
  \begin{aligned}
    \cyxy &= -\cxyy = \inv{\heta\hxi}\del{\heta}{\xi} &\qquad 
    \czxz &= -\cxzz = \inv{\hphi\hxi}\del{\hphi}{\xi} \\
    \cxyx &= -\cyxx = \inv{\hxi\heta}\del{\hxi}{\eta} &\qquad 
    \czyz &= -\cyzz = \inv{\hphi\heta}\del{\hphi}{\eta} \\
    \cxzx &= -\czxx = \inv{\hxi\hphi}\del{\hxi}{\phi} &\qquad
    \cyzy &= -\czyy = \inv{\heta\hphi}\del{\heta}{\phi}.
  \end{aligned} 
\end{equation}

\bibliographystyle{cpc}
\bibliography{numscheme}





\end{document}